\documentclass[aps,prb,reprint,superscriptaddress,longbibliography]{revtex4-2}
\usepackage[utf8]{inputenc}
\usepackage[T1]{fontenc}
\usepackage{lmodern}
\input{glyphtounicode}\pdfgentounicode=1
\usepackage{amsmath,amssymb,bm}
\usepackage{array}[=2016-10-06]
\usepackage{graphicx,xcolor,tabularx,array,fvextra}
\usepackage[colorlinks,allcolors=blue!50!black]{hyperref}
\providecommand{\tightlist}{\setlength{\itemsep}{0pt}\setlength{\parskip}{0pt}}
\newsavebox{\tblbox}\newcommand{\fitbox}[1]{\sbox{\tblbox}{#1}\ifdim\wd\tblbox>\linewidth\resizebox{\linewidth}{!}{\usebox{\tblbox}}\else\usebox{\tblbox}\fi}

\begin{document}
\title{Packing the Bragg sphere: complete three-dimensional photonic band gaps at the refractive index of polymers (1.55)}
\author{Alexander Itin}
\email{alexander.itin@ru.nl}
\homepage{https://AI4science.de}
\affiliation{Radboud University, Nijmegen, The Netherlands}
\affiliation{Independent researcher, Hamburg, Germany}
\date{\today}
\begin{abstract}Complete 3D photonic band gaps have so far been found only at refractive-index contrasts of about 1.9 and above, reached by rod-connected diamond lattice structures. Reciprocal-space design of quasiperiodic structures recently achieved considerable emission attenuation at much lower contrasts {[}L. Maiwald et al., Adv. Optical Mater. 10, 2100785 (2022){]}. Here we implement systematic search and reciprocal-space design of periodic 3D structures with complete band gaps, solving a discrete optimization problem on an integer lattice. A gap above a high band needs the strongest Fourier components of the dielectric close to the Bragg sphere of that band. Crystals with large cells, started from Fourier peaks at the vertices of a polyhedron on this sphere and optimized against densely refined band edges, open complete gaps down to an index contrast of 1.55, that of common polymers against air. A rational approximant of an icosidodecahedron gave the most efficient location of the Fourier peaks among the polyhedra tested. Numerical calculations confirm the gaps over a resolution series, with band edges searched over the whole Brillouin zone. The best designs are connected porous networks, suited to a microwave test with printed polymer. A rigorous bound, valid for any structure, does not exclude gaps at lower index, and the next rational approximants of the icosidodecahedron, whose Bragg zones match bands 576 and 2440 almost exactly, point to where gaps at even lower contrast may be sought.\end{abstract}
\maketitle
Over the past decades, photonic band gaps have been explored extensively in periodic, aperiodic and quasicrystalline dielectrics \cite{Yablonovitch1987,John1987,HoChanSoukoulis1990,YablonovitchGmitterLeung1991,MaldovanThomas2004,Man2005,DyachenkoMiklyaev2006,Dyachenko2007,Rechtsman2008,SigmundHougaard2008,FlorescuPRB2009,FlorescuPNAS2009,Men2014,CerjanFan2017,Chen2019,Cersonsky2021,Christensen2022,Maiwald2022,Saha2023,Siedentop2024,Zhang2024,Casiulis2025,Itin2025}. They underpin many ways of confining and steering light, and living organisms such as beetles produce them too \cite{VukusicSambles2003,Galusha2008}. Already in 1887 Rayleigh showed that waves in a periodically stratified medium are reflected in bands of frequency, and that such a stop band opens no matter how weak the stratification is \cite{Rayleigh1887}. It was noted that an atom inside such a structure radiates less \cite{Bykov1972}, and in 1987 Yablonovitch and John asked for the three-dimensional version: a crystal that forbids light of every direction and polarization in a band of frequencies, to inhibit spontaneous emission and to localize light \cite{Yablonovitch1987,John1987}. In three dimensions the stop bands of all directions must overlap, and that requires sufficient optical contrast between the constituent materials. The first complete gap was predicted for the diamond lattice \cite{HoChanSoukoulis1990} and shown at microwave frequencies in a block of high-index dielectric drilled along three directions \cite{YablonovitchGmitterLeung1991}.

Among ordinary dielectric structures, complete gaps have so far been verified only at refractive-index contrasts of about 1.9 or more. The threshold belongs to the diamond family \cite{HoChanSoukoulis1990,MaldovanThomas2004} and was confirmed in Sn--S--O inverse rod-connected diamond \cite{Chen2019}. Topology optimization of every voxel improved on the hand designs only slightly \cite{Men2014}, and the largest survey of crystal templates covered only high contrast and low bands \cite{Cersonsky2021}. Printable polymers and glasses have indices near 1.5 \cite{Gissibl2017}, so a complete gap there needs backfilling with a high-index material \cite{Chen2019} or printing in titania \cite{Zhang2024}. Whether 1.9 is a limit of three dimensions, or of the low bands where the search has taken place \cite{Christensen2022}, is still an open question.

At low contrast a gap is a Bragg effect, and design moves to reciprocal space: the Bragg planes of the strongest Fourier components should approximately enclose the free-photon states of the lowest n bands, like the Jones zones of metal physics \cite{Jones1934}. In two dimensions a ring of Bragg peaks does this. Optimized quasicrystals \cite{Rechtsman2008,FlorescuPRB2009}, hyperuniform patterns \cite{FlorescuPNAS2009}, quasicrystal interference fields \cite{DyachenkoMiklyaev2006} and gyromorphs \cite{Casiulis2025} all carry such a ring, and supercells markedly lowered the contrast that a complete 2D gap needs \cite{CerjanFan2017}.

In three dimensions, Maiwald et al. \cite{Maiwald2022} asked for the minimal contrast directly. For quasicrystals built from superposed gratings they estimated a finite complete gap at any contrast, in 2D and in 3D. In 3D simulations the emission is suppressed only partly, short of that prediction, and millimetre-scale polymer prints suppress microwave transmission \cite{Maiwald2022,Saha2023}. Cubic approximants of an icosahedral quasicrystal have large, nearly isotropic complete gaps over a wide range of permittivity \cite{Dyachenko2007}, stealthy hyperuniform networks show an isotropic gap at n \ensuremath{\approx} 2.1 in microwaves \cite{Siedentop2024}, and gyromorphs show one in a coupled-dipole model \cite{Casiulis2025}. A complete gap below 1.9 has thus been predicted but not established.

We look for the gap in crystals with large cells, where it is a band-structure statement that standard solvers can check at every wavevector. Three results follow. (i) Free-form level-set structures, optimized with exact gradients against densely refined band edges, open complete gaps below the diamond threshold. (ii) Their Fourier power gathers in a thin shell at the Bragg sphere, spread over many pairs of lattice vectors, and no single Jones zone accounts for it; the best designs grow from peaks placed at the vertices of a polyhedron on that sphere. (iii) An independent solver confirms complete gaps on all three cubic lattices, down to the index of common polymers on the body-centred-cubic lattice. A rigorous empty-lattice bound, valid for any structure, lies far below these contrasts, so gaps at lower index are not ruled out. The Letter is self-contained; the Supplemental Material (SM \cite{SM}) adds derivations, implementation details and the complete verification records.

A two-phase structure \ensuremath{\varepsilon}(r) = 1 + (c \ensuremath{-} 1)\ensuremath{\theta}(r) is defined in a primitive cell of unit volume of the sc, bcc or fcc lattice. The lattice fixes only the translations: no point symmetry is imposed, and the final designs have none; \ensuremath{\theta} is the indicator of the dense phase, f its filling fraction, and the index contrast is \ensuremath{\surd}c. In the empty lattice, counting both polarizations, n free-photon states fill a sphere of radius \(k_{n}\), the characteristic wavevector of band n:

\begin{equation}
n = \frac{8\pi}{3}\,k_n^{3}, \qquad R \equiv 2k_n = 2\left(\frac{3n}{8\pi}\right)^{1/3}
\end{equation}

with wavevectors in units of the reciprocal lattice (no factor 2\ensuremath{\pi}). The Fourier component \(\hat{\theta}\)(G) couples free photons whose wavevectors differ by G. Gaps above band n need strong components on the Bragg sphere \textbar G\textbar{} \ensuremath{\approx} R. A complete gap lies between the highest frequency of band n and the lowest of band n + 1, both taken over the whole Brillouin zone and not only on its symmetry lines; its relative width is

\begin{equation}
\begin{gathered}\Delta = \frac{2(\omega_+ - \omega_-)}{\omega_+ + \omega_-}, \\ \omega_- = \max_{\mathbf k}\,\omega_n(\mathbf k), \qquad \omega_+ = \min_{\mathbf k}\,\omega_{n+1}(\mathbf k).\end{gathered}
\end{equation}

Min--max monotonicity in \ensuremath{\varepsilon} gives a rigorous bound for any structure with 1 \ensuremath{\leq} \ensuremath{\varepsilon} \ensuremath{\leq} c: a gap of relative width \ensuremath{\Delta} above band n requires \ensuremath{\surd}c \ensuremath{\geq} \(R_{n}\) (1 + \ensuremath{\Delta}/2)/(1 \ensuremath{-} \ensuremath{\Delta}/2), where \(R_{n}\) is the ratio of the maximum of band n to the minimum of band n + 1 in the empty lattice (SM \cite{SM}, S7). The bound \(c_{B}\)(n) for a 0.3 \% gap depends strongly on n: sc 88 gives 1.132, bcc 136 gives 1.112, fcc 190 gives 1.105, while sc 56 gives 1.46. We design only at local minima of \(c_{B}\).

The Bragg planes of the strongest peaks bound a polyhedron, the analogue of a Jones zone, which approximately represents the states below the gap. In a sharp-zone model a gap above band n would require it to hold n/2 plane waves at every k, so to have volume n/2 (area n in 2D TM) and to tile reciprocal space n/2 times under lattice translations \cite{GravinRobinsShiryaev2012}. At finite contrast the modes mix plane waves across the zone boundary, so neither condition is necessary; the zone volume and the spread of its translated lattice-point count are diagnostics of a start. In 2D the strongest ring peaks reproduce the area n. In 3D the strong peaks of designs grown from evenly filled shells do not form such a zone. The zone of the best polyhedral start holds exactly n/2 = 68 plane waves at 94 \% of 200 000 random Bloch vectors (range 66--68, rms deviation 0.27). For every other start tried the rms deviation is 2.5--7.6, also where the volume matches n, and all of them end at higher contrast (SM \cite{SM}, S12).

A first-order count balances the Bragg opening of N peaks on the sphere, which falls as N\(^{\ensuremath{-}1/2}\), against the uncovered directions between them, which shrink as 1/N. Its optimum lies where both terms are second order in the contrast, like the couplings it drops, so it fixes neither N nor the gap; this holds for the estimate of \cite{Maiwald2022} and for our own version. The designs use a share of the shell instead: their effective number of shell pairs grows with the sphere, from about 18 at R = 3.4 to 100--115 at R \ensuremath{\approx} 6, a third of the pairs the lattice offers, and the weak peaks count: in a coarse model, dropping the 63 weakest of the 313 pairs of the design at index 1.654, 1.6 \% of the power of its generating series, at fixed filling, closes its gap (SM \cite{SM}, S9). The optimal number at given R remains to be found.

The structure is the smoothed level set \ensuremath{\theta}(r) = \(S_{\ensuremath{\tau}}\){[}\ensuremath{\varphi}(r){]} of a real Fourier series \ensuremath{\varphi}(r) = \(\ensuremath{\Sigma}_{G}\) \(a_{G}\) cos(2\ensuremath{\pi}G\ensuremath{\cdot}r + \(\ensuremath{\alpha}_{G}\)), with 218--313 coefficient pairs (0.15 \ensuremath{\leq} \textbar G\textbar/2\(k_{n}\) \ensuremath{\leq} 1.06). Bands come from a plane-wave solver for the H field, with plane waves up to \textbar k + G\textbar{} \ensuremath{\leq} 2\(k_{n}\) and Ho\textquotesingle s inverse rule \cite{HoChanSoukoulis1990}. Hellmann--Feynman derivatives of the band edges give exact gradients in all coefficients. L-BFGS maximizes the gap above band n while the contrast is lowered step by step, c \ensuremath{\rightarrow} 1 + (c \ensuremath{-} 1) \ensuremath{\times} 0.96, from a start at c \ensuremath{\approx} 4--4.5. Starts: random shells, shells weighted by Riesz charges (amplitudes \ensuremath{\surd}\(q_{j}\), with charges \(q_{j}\) \ensuremath{\geq} 0 on the shell directions \ensuremath{\pm}\(G_{j}\)/\textbar{}\(G_{j}\)\textbar{} that come closest to equilibrium under logarithmic repulsion; slightly better), sums of cylindrical waves (below), or polyhedral peak sets (best). The threshold \ensuremath{\tau} sharpens to a binary structure at the end.

A design optimized against a fixed k-set learns to push band extrema between its points; our first apparent gaps at index 1.58--1.59 closed on finer grids.

Every design is optimized against refined band edges: in each round a scan on two k grids and 200 random points of the Brillouin zone, followed by Nelder--Mead refinement of the four highest maxima of band n and the four lowest minima of band n + 1, adds points to the working k-set. The final design is checked on the two grids, 400 random points and twelve plus twelve refinements; the contrast at which this plane-wave gap reaches 0.3 \% is \(c_{03}\).

A gap is claimed only from the independent solver MIT Photonic Bands (MPB) \cite{JohnsonJoannopoulos2001} on the binary structure, on a contrast ladder starting at 0.98 \(c_{03}\): a scan at resolution 40 on two k grids and the plane-wave extremal set, refinement of the band edges from the ten highest and ten lowest scan points, and the edge set recomputed at resolutions 48, 64 and 80 and extrapolated linearly in 1/res\(^{2}\). A claim requires the refined gap to exceed an empirical acceptance margin, three times the sum of the refinement loss and the spread between extrapolations linear in 1/res\(^{2}\) and in 1/res; smaller positive gaps are candidates. The main designs pass two further checks: a global search from about 1600 starts, with MPB refinement of every extremum near the band edges, and a numerical covering certificate that bounds the bands of the finite plane-wave model over the whole Brillouin zone. The certificate concerns that model, not Maxwell\textquotesingle s equations. All gaps below are refined-edge values; they grow with resolution in every case. Implementation and validation details are given in SM \cite{SM}, S2--S5.

MPB confirms complete gaps in eighteen designs on the three cubic lattices; Table I lists the best of each kind. The lowest contrast belongs to a bcc design above band 136, grown from Fourier peaks at the 30 vertices of an approximate icosidodecahedron on the Bragg sphere (Fig. 1) and improved by two rounds of perturbed restarts. Its gap at index contrast 1.547 is 0.25 \% on the refined band edges and grows to 0.31 \% at resolution 80 (Fig. 2); the global band-edge search reproduces it, and a numerical covering certificate holds for its plane-wave model. That is 19 \% below the diamond threshold in index and 47 \% below it in c \ensuremath{-} 1. Designs grown from evenly filled shells stop at index 1.584 on bcc and 1.677 on fcc, and a polyhedral start reaches 1.681 on sc. Eight further designs have positive gaps at every resolution but fail the acceptance margin; they count as candidates.

\begin{table*}[tb]
\caption{The best MPB-confirmed design of each kind. Gaps are \ensuremath{\Delta}\ensuremath{\omega}/\(\ensuremath{\omega}_{\mathrm{mid}}\) on the refined band-edge set (up to 12 k-points), with n + 10 bands and tolerance 10\(^{-8}\); the extrapolation is linear in 1/res\(^{2}\) over res 48--80. Res 40: refined edges (*: from the global search). Polyh.: polyhedral start; a round is one round of perturbed restarts. All claims, candidates and gap openings are in SM \cite{SM}, S6.}
\centering\small
\fitbox{\begin{tabular}{lccc}
\hline\hline
Design & c (index) & Gap (\%), res 40 / 48 / 64 / 80 & Res \ensuremath{\rightarrow} \ensuremath{\infty} \\
\hline
bcc 136 polyh., 2 rounds & 2.3925 (1.547) & 0.252 / 0.279 / 0.301 / 0.308 & 0.326 \\
bcc 136 polyh., 1 round* & 2.4487 (1.565) & 0.284 / 0.289 / 0.302 / 0.307 & 0.318 \\
bcc 136 Riesz, 2 rounds & 2.5084 (1.584) & 0.250 / 0.286 / 0.327 / 0.343 & 0.377 \\
bcc 136 Riesz & 2.7343 (1.654) & 0.428 / 0.449 / 0.464 / 0.475 & 0.488 \\
fcc 118 Riesz, 2 rounds & 2.8108 (1.677) & 0.409 / 0.436 / 0.465 / 0.479 & 0.503 \\
sc 88 polyh., 1 round & 2.8272 (1.681) & 0.424 / 0.448 / 0.478 / 0.491 & 0.515 \\
sc 88 random & 3.0926 (1.759) & 0.304 / 0.313 / 0.332 / 0.343 & 0.358 \\
\hline\hline
\end{tabular}}
\end{table*}

The lowest contrasts come from starting the optimization with Fourier peaks at the vertices of a polyhedron on the Bragg sphere, instead of a shell filled evenly (Fig. 1). On bcc 136 the 30 lattice vectors closest to an icosidodecahedron work best. The structure is real, so peaks come in pairs \ensuremath{\pm}G, and spreading them evenly is the Tammes problem \cite{MusinTarasov2015} for lines through the origin \cite{ConwaySloane1993,ConwayHardinSloane1996,HardinSloaneSmith}. Restricted to lattice vectors, the problem becomes discrete: we combine a search over start sets with continuous optimization of all Fourier coefficients. Icosahedral solids have the roundest Bragg polyhedra but no lattice form, so their vertices need rational approximants of the golden ratio \ensuremath{\varphi}. The bcc vectors \{600\} + \{532\}c (c: cyclic permutations only), built from the Fibonacci numbers 2, 3 and 5, point within 1.2\textdegree{} of the 30 vertices of an icosidodecahedron (Fig. 1b); their lengths, 6 and \ensuremath{\surd}38, differ by 2.7 \%. Their Bragg planes bound an approximate rhombic triacontahedron of doubled volume 2038/15 = 135.87, against the 136 bands below the gap. For this bcc band-136 structure the bare level set has no gap, but the optimized design passes the claim test at index 1.572 as it stands; after one round of perturbed restarts its gap holds down to 1.555, and a second round reaches 1.547. The same rounds stopped the Riesz-start design at 1.584. Starts on other polyhedra (Fig. 3), and on the other lattices, end at higher index.

This match of zone and band is arithmetic. For consecutive Fibonacci numbers r \textless{} q \textless{} p, with q = \(F_{m}\), the 30 bcc vectors \{2q 0 0\} + \{pqr\}c bound a zone of doubled volume

\begin{equation}
N_m = 4F_{3m-3} - \frac{2F_{m-3}}{F_m F_{m+1}}
\end{equation}

that is 136 \ensuremath{-} 2/15, 576 \ensuremath{-} 1/20 and 2440 \ensuremath{-} 1/26 for m = 4, 5 and 6 (proof in SM \cite{SM}, S12); whether band 576 does better is an open question.

Closer approximants come nearer to the icosidodecahedron in both radius and direction, but its holes remain: the widest, on the five-fold axes, have half-angle arctan(1/\ensuremath{\varphi}) = 31.7\textdegree{}. A wave there meets the nearest Bragg plane 17.6 \% beyond the face centres, and that plane couples its weaker polarization with a factor \textbar cos 2\ensuremath{\theta}\textbar{} = 1/\ensuremath{\surd}5. Thirty peaks alone thus leave the five-fold directions weakly coupled, in line with the missing gap of the bare start.

All three bands sit at deep local minima of the empty-lattice bound (Fig. 4); bcc 136 has the lowest bound of the three and is the only one of these cells with a close approximant of the icosidodecahedron.

\begin{figure}[tb]
\includegraphics[width=\linewidth]{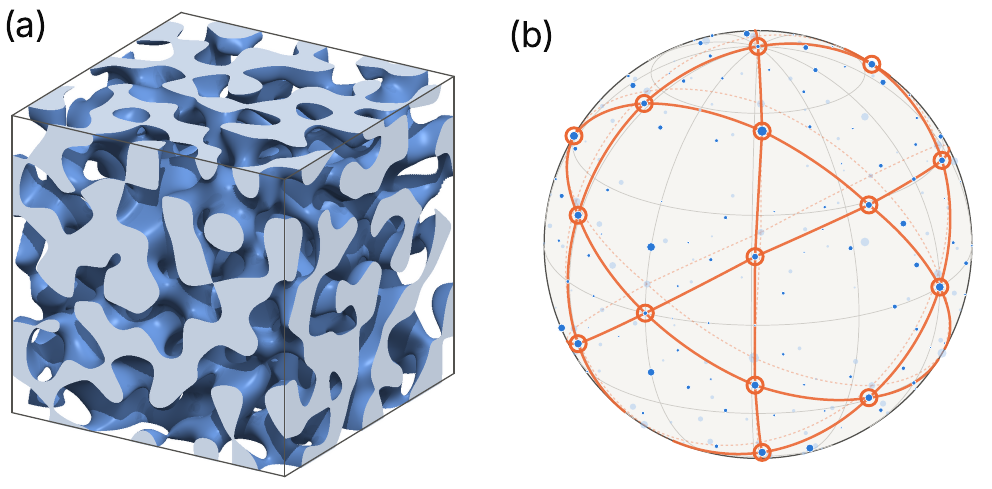}
\caption{The design at index 1.547. (a) Its polymer phase (37 \% of the volume) in one conventional cubic cell of bcc 136, of edge 2.6 \ensuremath{\lambda}\(_{0}\) at mid-gap; the light faces are cuts through the polymer at the cell boundary. Polymer and air each form one connected network. (b) Directions of its Fourier peaks \(\hat{\theta}\)(G) in the shell 0.85--1.12 \ensuremath{\times} 2\(k_{n}\) (blue, area \ensuremath{\propto} \textbar{}\(\hat{\theta}\)\textbar{}\(^{2}\); pale: the far half of the sphere), with the 30 vectors of the icosidodecahedral start \{600\} + \{532\} (orange rings) and the edges of the icosidodecahedron; seen from the same direction as (a). The start vectors still carry the strongest peaks: the strongest of the 446 shell vectors and 26 of the strongest 40.}
\label{fig:1}
\end{figure}

\begin{figure*}[tb]
\includegraphics[width=\linewidth]{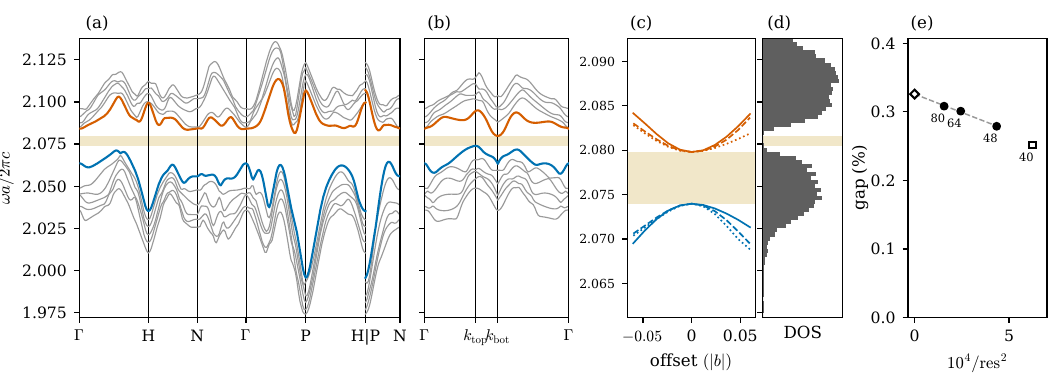}
\caption{Band structure of the bcc 136 design at c = 2.3925 (index 1.547). (a) Bands n \ensuremath{-} 5 \ldots{} n + 5 along \ensuremath{\Gamma}--H--N--\ensuremath{\Gamma}--P--H\textbar P--N; bands n (blue) and n + 1 (orange) have their extrema off these lines, so the gap on this path overstates the complete gap. (b) The path \ensuremath{\Gamma} \ensuremath{\rightarrow} \(k_{\mathrm{top}}\) \ensuremath{\rightarrow} \(k_{\mathrm{bottom}}\) \ensuremath{\rightarrow} \ensuremath{\Gamma} through the refined band edges and (c) six cuts through them, on an expanded scale. (d) Density of states on a 12\(^{3}\) grid; the grid misses the minimum of band n + 1, so its gap is wider. (e) Gap against 1/res\(^{2}\): refined edges at resolution 40 (square), the edge set at 48, 64 and 80 (dots), extrapolation (diamond). Shaded: the complete gap on the edge path at resolution 48.}
\label{fig:2}
\end{figure*}

The optimized structures are packings of the Bragg sphere, not of real-space objects. The best designs put about 18 strong pairs of peaks on a thin shell at 0.93--1.0 \ensuremath{\times} 2\(k_{n}\), spread evenly over directions. The design at index 1.547 shows where such a shell can come from: it grew from 30 peaks at the vertices of an approximate icosidodecahedron, and these still carry its strongest peaks (Fig. 1). They hold 36 \% of the Fourier power in the shell of Fig. 1(b). The next 10 \% sits in the 6 pairs \{602\}c, one in each five-fold hole, 12.5\textdegree{} off its axis and at 0.991 \ensuremath{\times} 2\(k_{n}\); IDIC\textquotesingle s icosahedron at the hole centres, at 0.914 \ensuremath{\times} 2\(k_{n}\), is removed (Fig. 4; SM \cite{SM}, S12).

\begin{figure}[tb]
\includegraphics[width=\linewidth]{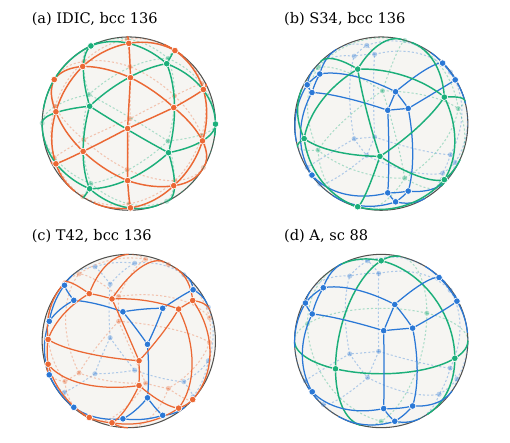}
\caption{Other polyhedral starts (SM \cite{SM}, S12): peak directions on the Bragg sphere, one colour and edge set per cubic orbit; pale: far half. (a) IDIC: the icosidodecahedron of Fig. 1(b) with its own icosahedron \{350\}c; (b) S34: \{433\} with the same icosahedron \{350\}c; (c) T42: \{442\} and \{541\}c; (d) A: the octahedron \{400\} with \{322\}, on sc 88. Best claims: IDIC at index 1.594, T42 at 1.649, S34 at 1.661 and A at 1.681.}
\label{fig:3}
\end{figure}

The linear response of the gap to each Fourier coefficient shows the division of labour. Scaling all shell peaks (0.85--1.05 \ensuremath{\times} 2\(k_{n}\)) by 1 \% widens the gap by 7--9 \% of itself in the Riesz-start designs. The inner peaks are as sensitive, but their contributions largely cancel (SM \cite{SM}, S9).

Sums of cylindrical waves test the coverage picture directly. A cylindrical wave J\(_{0}\)(K\ensuremath{\rho}) about an axis u has its Fourier weight on the great circle of the Bragg sphere perpendicular to u. Axes along the 6 icosahedral five-fold directions cover the sphere most evenly and do best (bcc 136: \(c_{03}\) = 2.805, close to 2.790 for the Riesz start). The 3 cube axes, which leave the sphere poorly covered, do worst (bcc 136: \(c_{03}\) = 3.20; fcc 118: no gap; SM \cite{SM}, S8).

The same method improves 2D designs. For a 0.3 \% TM gap it needs 8--21 \% less c \ensuremath{-} 1 than the best equal-rod structures, in MPB checks at four bands; these are single-polarization gaps, unlike the complete gaps of \cite{CerjanFan2017}. There too, strong peaks on the Bragg circle work with weaker inner ones. The triangular n = 12 rod record carries a regular 12-gon of peaks on that circle, and its strong inner peaks, phase-locked differences of ring vectors, couple the states at the zone\textquotesingle s vertices; starts from the ring alone reach the rod records but do not pass them (SM \cite{SM}, S13).

A complete 3D gap at the index of common polymers is now established numerically, by MPB with a global band-edge search and by a numerical covering certificate of its plane-wave model, and the reciprocal-space picture points to higher bands at minima of the empty-lattice bound, with more peaks on a larger Bragg sphere.

At high contrast, several geometric ideas have been proposed for designing photonic crystals with large band gaps \cite{SigmundHougaard2008,Itin2025}, among them optimal packings, one of the oldest problems of mathematics \cite{Weaire1999,AsteWeaire2000,MusinNikitenko2016,ConnellyDickinson2014}: rods at the centres of optimal toric disk packings, with a single parameter, rival topology optimization with hundreds of parameters \cite{Itin2025}. Packing Fourier peaks on the Bragg sphere was already used for quasiperiodic structures \cite{Maiwald2022}; here we find other, more involved geometric recipes for placing Fourier peaks on the integer lattice of reciprocal space.

Index 1.5--1.6 covers common polymers and glasses against air, and the polyhedral designs now reach the middle of that range. Other polyhedra do not go further: every other family so far ends at index 1.594 or higher (Fig. 3). Nor do the filling and the number of peaks: with the support restricted to a few dozen well-spread shell pairs no gap opens at these contrasts, and with the full support the filling returns to 0.33--0.37 from any start (SM \cite{SM}, S9). A one-line estimate gives this value: for a thresholded Gaussian field the Bragg strength of a shell peak, (c \ensuremath{-} 1)\ensuremath{\varphi}(t)/{[}1 + (c \ensuremath{-} 1)f{]} with f = 1 \ensuremath{-} \ensuremath{\Phi}(t) (\ensuremath{\varphi}, \ensuremath{\Phi}: normal density and distribution), is largest at f = 0.36 for c = 2.39. What remains is larger cells, which have not yet improved on n = 136 (Fig. 4), so whether they help remains to be seen.

\begin{figure}[tb]
\includegraphics[width=\linewidth]{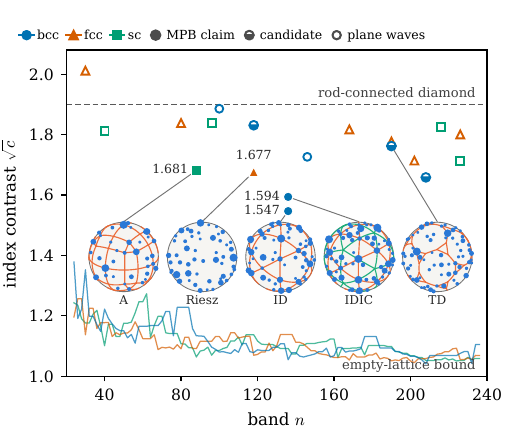}
\caption{Index contrast \ensuremath{\surd}c of complete gaps above band n. Lines: the empty-lattice bound; below its line no structure has a 0.3 \% gap above band n. Filled markers: MPB claims, labelled with the index; half-filled: MPB candidates; open: the best plane-wave design at other bands. At bcc 136 the best design lies 12.4 times above the bound in c \ensuremath{-} 1. Insets: Fourier peaks of five designs on the Bragg sphere (area \ensuremath{\propto} \textbar{}\(\hat{\theta}\)(G)\textbar{}\(^{2}\), near half) and their starting polyhedra (Fig. 3; TD, truncated dodecahedron); optimization drops some vertex peaks (e.g., IDIC\textquotesingle s icosahedron) and adds others.}
\label{fig:4}
\end{figure}

The result establishes a complete gap at this contrast but is not yet a device; the experiment within reach is a printed sample at microwave frequencies, like the polymer prints of quasicrystals \cite{Maiwald2022,Man2005}. The design needs no special material: its index is that of common printing filaments such as PLA and HIPS (\ensuremath{\varepsilon} \ensuremath{\approx} 2.4--2.5 at 1--67 GHz) \cite{PerezEscribano2021}, and polymer and air each form one connected network, so it prints as a single piece (Fig. 1a). At tens of gigahertz the cell measures centimetres and the struts millimetres, the feature size of these prints. The gap is narrow, but the two bands that bound it are nearly flat, so the field inside the gap decays within a few cells, and a block of about ten cells should show a deep stop band (estimated for the design at index 1.654; SM \cite{SM}, S11). What remains demanding is uniformity: in that design a drift of one or two percent in filling or index moves the gap by its own width, and the filament\textquotesingle s index must be measured first and the design retuned to it. The min--max argument also guarantees that a gap of relative width \ensuremath{\Delta} survives any cell-periodic error of the permittivity below \ensuremath{\Delta}/(1 + \ensuremath{\Delta}\(^{2}\)/4) in relative size, about 0.3 \% here; displaced interfaces are not covered. At optical wavelengths two-photon polymerization writes such shapes \cite{Deubel2004,Hu2026}, but not yet with that uniformity.

The structures are crystals with large cells whose Fourier weight imitates an isotropic shell, as gyromorphs and distributed quasicrystals do with true disorder or quasiperiodicity. Periodicity costs some isotropy, but it makes the gap verifiable in a band structure.

All designs, more than 80 000 3D structures with bands, and the MPB claim data are kept for surrogate models and ML-backed design, continuing Refs. \cite{Nikulin2022,Schurov2024}.

\begin{acknowledgments}
We thank Profs. M. Titov, M. Katsnelson and A. Bagrov for hospitality at Radboud University, whose computing cluster ran most of the calculations; colleagues at Hamburg University of Technology, Profs. A. Petrov and M. Eich, for suggesting the problem and early discussions on reciprocal-space design and printed photonic structures. The author used AI assistants (Claude Opus 5.5 and Fable 5.2, Anthropic; Astra 6, OpenAI) for programming, running and analysing numerical calculations, reviewing drafts, and editing the text (the research questions, the design principle and the interpretation of the results are the author\textquotesingle s own; the author has checked the derivations, code and reported numbers and takes full responsibility for the content).

\end{acknowledgments}
\section*{Data availability}
The designs, their structures and band diagrams, and scripts that rebuild the structures and check their gaps with MPB are available from the author upon reasonable request, as are the other code and data (SM \cite{SM}, S10).
\bibliography{refs}

\begin{thebibliography}{47}%
\makeatletter
\providecommand \@ifxundefined [1]{%
 \@ifx{#1\undefined}
}%
\providecommand \@ifnum [1]{%
 \ifnum #1\expandafter \@firstoftwo
 \else \expandafter \@secondoftwo
 \fi
}%
\providecommand \@ifx [1]{%
 \ifx #1\expandafter \@firstoftwo
 \else \expandafter \@secondoftwo
 \fi
}%
\providecommand \natexlab [1]{#1}%
\providecommand \enquote  [1]{``#1''}%
\providecommand \bibnamefont  [1]{#1}%
\providecommand \bibfnamefont [1]{#1}%
\providecommand \citenamefont [1]{#1}%
\providecommand \href@noop [0]{\@secondoftwo}%
\providecommand \href [0]{\begingroup \@sanitize@url \@href}%
\providecommand \@href[1]{\@@startlink{#1}\@@href}%
\providecommand \@@href[1]{\endgroup#1\@@endlink}%
\providecommand \@sanitize@url [0]{\catcode `\\12\catcode `\$12\catcode
  `\&12\catcode `\#12\catcode `\^12\catcode `\_12\catcode `\%12\relax}%
\providecommand \@@startlink[1]{}%
\providecommand \@@endlink[0]{}%
\providecommand \url  [0]{\begingroup\@sanitize@url \@url }%
\providecommand \@url [1]{\endgroup\@href {#1}{\urlprefix }}%
\providecommand \urlprefix  [0]{URL }%
\providecommand \Eprint [0]{\href }%
\providecommand \doibase [0]{https://doi.org/}%
\providecommand \selectlanguage [0]{\@gobble}%
\providecommand \bibinfo  [0]{\@secondoftwo}%
\providecommand \bibfield  [0]{\@secondoftwo}%
\providecommand \translation [1]{[#1]}%
\providecommand \BibitemOpen [0]{}%
\providecommand \bibitemStop [0]{}%
\providecommand \bibitemNoStop [0]{.\EOS\space}%
\providecommand \EOS [0]{\spacefactor3000\relax}%
\providecommand \BibitemShut  [1]{\csname bibitem#1\endcsname}%
\let\auto@bib@innerbib\@empty
\bibitem [{\citenamefont {Yablonovitch}(1987)}]{Yablonovitch1987}%
  \BibitemOpen
  \bibfield  {author} {\bibinfo {author} {\bibfnamefont {E.}~\bibnamefont
  {Yablonovitch}},\ }\bibfield  {title} {\bibinfo {title} {Inhibited
  spontaneous emission in solid-state physics and electronics},\ }\href@noop {}
  {\bibfield  {journal} {\bibinfo  {journal} {Phys. Rev. Lett.}\ }\textbf
  {\bibinfo {volume} {58}},\ \bibinfo {pages} {2059} (\bibinfo {year}
  {1987})}\BibitemShut {NoStop}%
\bibitem [{\citenamefont {John}(1987)}]{John1987}%
  \BibitemOpen
  \bibfield  {author} {\bibinfo {author} {\bibfnamefont {S.}~\bibnamefont
  {John}},\ }\bibfield  {title} {\bibinfo {title} {Strong localization of
  photons in certain disordered dielectric superlattices},\ }\href@noop {}
  {\bibfield  {journal} {\bibinfo  {journal} {Phys. Rev. Lett.}\ }\textbf
  {\bibinfo {volume} {58}},\ \bibinfo {pages} {2486} (\bibinfo {year}
  {1987})}\BibitemShut {NoStop}%
\bibitem [{\citenamefont {Ho}\ \emph {et~al.}(1990)\citenamefont {Ho},
  \citenamefont {Chan},\ and\ \citenamefont {Soukoulis}}]{HoChanSoukoulis1990}%
  \BibitemOpen
  \bibfield  {author} {\bibinfo {author} {\bibfnamefont {K.~M.}\ \bibnamefont
  {Ho}}, \bibinfo {author} {\bibfnamefont {C.~T.}\ \bibnamefont {Chan}},\ and\
  \bibinfo {author} {\bibfnamefont {C.~M.}\ \bibnamefont {Soukoulis}},\
  }\bibfield  {title} {\bibinfo {title} {Existence of a photonic gap in
  periodic dielectric structures},\ }\href
  {https://doi.org/10.1103/PhysRevLett.65.3152} {\bibfield  {journal} {\bibinfo
   {journal} {Phys. Rev. Lett.}\ }\textbf {\bibinfo {volume} {65}},\ \bibinfo
  {pages} {3152} (\bibinfo {year} {1990})}\BibitemShut {NoStop}%
\bibitem [{\citenamefont {Yablonovitch}\ \emph {et~al.}(1991)\citenamefont
  {Yablonovitch}, \citenamefont {Gmitter},\ and\ \citenamefont
  {Leung}}]{YablonovitchGmitterLeung1991}%
  \BibitemOpen
  \bibfield  {author} {\bibinfo {author} {\bibfnamefont {E.}~\bibnamefont
  {Yablonovitch}}, \bibinfo {author} {\bibfnamefont {T.~J.}\ \bibnamefont
  {Gmitter}},\ and\ \bibinfo {author} {\bibfnamefont {K.~M.}\ \bibnamefont
  {Leung}},\ }\bibfield  {title} {\bibinfo {title} {Photonic band structure:
  The face-centered-cubic case employing nonspherical atoms},\ }\href
  {https://doi.org/10.1103/PhysRevLett.67.2295} {\bibfield  {journal} {\bibinfo
   {journal} {Phys. Rev. Lett.}\ }\textbf {\bibinfo {volume} {67}},\ \bibinfo
  {pages} {2295} (\bibinfo {year} {1991})}\BibitemShut {NoStop}%
\bibitem [{\citenamefont {Maldovan}\ and\ \citenamefont
  {Thomas}(2004)}]{MaldovanThomas2004}%
  \BibitemOpen
  \bibfield  {author} {\bibinfo {author} {\bibfnamefont {M.}~\bibnamefont
  {Maldovan}}\ and\ \bibinfo {author} {\bibfnamefont {E.~L.}\ \bibnamefont
  {Thomas}},\ }\bibfield  {title} {\bibinfo {title} {Diamond-structured
  photonic crystals},\ }\href {https://doi.org/10.1038/nmat1201} {\bibfield
  {journal} {\bibinfo  {journal} {Nat. Mater.}\ }\textbf {\bibinfo {volume}
  {3}},\ \bibinfo {pages} {593} (\bibinfo {year} {2004})}\BibitemShut {NoStop}%
\bibitem [{\citenamefont {Man}\ \emph {et~al.}(2005)\citenamefont {Man},
  \citenamefont {Megens}, \citenamefont {Steinhardt},\ and\ \citenamefont
  {Chaikin}}]{Man2005}%
  \BibitemOpen
  \bibfield  {author} {\bibinfo {author} {\bibfnamefont {W.}~\bibnamefont
  {Man}}, \bibinfo {author} {\bibfnamefont {M.}~\bibnamefont {Megens}},
  \bibinfo {author} {\bibfnamefont {P.~J.}\ \bibnamefont {Steinhardt}},\ and\
  \bibinfo {author} {\bibfnamefont {P.~M.}\ \bibnamefont {Chaikin}},\
  }\bibfield  {title} {\bibinfo {title} {Experimental measurement of the
  photonic properties of icosahedral quasicrystals},\ }\href@noop {} {\bibfield
   {journal} {\bibinfo  {journal} {Nature}\ }\textbf {\bibinfo {volume}
  {436}},\ \bibinfo {pages} {993} (\bibinfo {year} {2005})}\BibitemShut
  {NoStop}%
\bibitem [{\citenamefont {Dyachenko}\ and\ \citenamefont
  {Miklyaev}(2006)}]{DyachenkoMiklyaev2006}%
  \BibitemOpen
  \bibfield  {author} {\bibinfo {author} {\bibfnamefont {P.~N.}\ \bibnamefont
  {Dyachenko}}\ and\ \bibinfo {author} {\bibfnamefont {Y.~V.}\ \bibnamefont
  {Miklyaev}},\ }\bibfield  {title} {\bibinfo {title} {Band structure
  calculations of {2D} photonic pseudoquasicrystals obtainable by holographic
  lithography},\ }in\ \href@noop {} {\emph {\bibinfo {booktitle} {Proc.
  SPIE}}},\ Vol.\ \bibinfo {volume} {6182}\ (\bibinfo {year} {2006})\ p.\
  \bibinfo {pages} {61822I}\BibitemShut {NoStop}%
\bibitem [{\citenamefont {Dyachenko}\ \emph {et~al.}(2007)\citenamefont
  {Dyachenko}, \citenamefont {Miklyaev},\ and\ \citenamefont
  {Dmitrienko}}]{Dyachenko2007}%
  \BibitemOpen
  \bibfield  {author} {\bibinfo {author} {\bibfnamefont {P.~N.}\ \bibnamefont
  {Dyachenko}}, \bibinfo {author} {\bibfnamefont {Y.~V.}\ \bibnamefont
  {Miklyaev}},\ and\ \bibinfo {author} {\bibfnamefont {V.~E.}\ \bibnamefont
  {Dmitrienko}},\ }\bibfield  {title} {\bibinfo {title} {Three-dimensional
  photonic quasicrystal with a complete band gap},\ }\href
  {https://doi.org/10.1134/S0021364007160047} {\bibfield  {journal} {\bibinfo
  {journal} {JETP Lett.}\ }\textbf {\bibinfo {volume} {86}},\ \bibinfo {pages}
  {240} (\bibinfo {year} {2007})}\BibitemShut {NoStop}%
\bibitem [{\citenamefont {Rechtsman}\ \emph {et~al.}(2008)\citenamefont
  {Rechtsman}, \citenamefont {Jeong}, \citenamefont {Chaikin}, \citenamefont
  {Torquato},\ and\ \citenamefont {Steinhardt}}]{Rechtsman2008}%
  \BibitemOpen
  \bibfield  {author} {\bibinfo {author} {\bibfnamefont {M.~C.}\ \bibnamefont
  {Rechtsman}}, \bibinfo {author} {\bibfnamefont {H.-C.}\ \bibnamefont
  {Jeong}}, \bibinfo {author} {\bibfnamefont {P.~M.}\ \bibnamefont {Chaikin}},
  \bibinfo {author} {\bibfnamefont {S.}~\bibnamefont {Torquato}},\ and\
  \bibinfo {author} {\bibfnamefont {P.~J.}\ \bibnamefont {Steinhardt}},\
  }\bibfield  {title} {\bibinfo {title} {Optimized structures for photonic
  quasicrystals},\ }\href@noop {} {\bibfield  {journal} {\bibinfo  {journal}
  {Phys. Rev. Lett.}\ }\textbf {\bibinfo {volume} {101}},\ \bibinfo {pages}
  {073902} (\bibinfo {year} {2008})}\BibitemShut {NoStop}%
\bibitem [{\citenamefont {Sigmund}\ and\ \citenamefont
  {Hougaard}(2008)}]{SigmundHougaard2008}%
  \BibitemOpen
  \bibfield  {author} {\bibinfo {author} {\bibfnamefont {O.}~\bibnamefont
  {Sigmund}}\ and\ \bibinfo {author} {\bibfnamefont {K.}~\bibnamefont
  {Hougaard}},\ }\bibfield  {title} {\bibinfo {title} {Geometric properties of
  optimal photonic crystals},\ }\href@noop {} {\bibfield  {journal} {\bibinfo
  {journal} {Phys. Rev. Lett.}\ }\textbf {\bibinfo {volume} {100}},\ \bibinfo
  {pages} {153904} (\bibinfo {year} {2008})}\BibitemShut {NoStop}%
\bibitem [{\citenamefont {Florescu}\ \emph
  {et~al.}(2009{\natexlab{a}})\citenamefont {Florescu}, \citenamefont
  {Torquato},\ and\ \citenamefont {Steinhardt}}]{FlorescuPRB2009}%
  \BibitemOpen
  \bibfield  {author} {\bibinfo {author} {\bibfnamefont {M.}~\bibnamefont
  {Florescu}}, \bibinfo {author} {\bibfnamefont {S.}~\bibnamefont {Torquato}},\
  and\ \bibinfo {author} {\bibfnamefont {P.~J.}\ \bibnamefont {Steinhardt}},\
  }\bibfield  {title} {\bibinfo {title} {Complete band gaps in two-dimensional
  photonic quasicrystals},\ }\href@noop {} {\bibfield  {journal} {\bibinfo
  {journal} {Phys. Rev. B}\ }\textbf {\bibinfo {volume} {80}},\ \bibinfo
  {pages} {155112} (\bibinfo {year} {2009}{\natexlab{a}})}\BibitemShut
  {NoStop}%
\bibitem [{\citenamefont {Florescu}\ \emph
  {et~al.}(2009{\natexlab{b}})\citenamefont {Florescu}, \citenamefont
  {Torquato},\ and\ \citenamefont {Steinhardt}}]{FlorescuPNAS2009}%
  \BibitemOpen
  \bibfield  {author} {\bibinfo {author} {\bibfnamefont {M.}~\bibnamefont
  {Florescu}}, \bibinfo {author} {\bibfnamefont {S.}~\bibnamefont {Torquato}},\
  and\ \bibinfo {author} {\bibfnamefont {P.~J.}\ \bibnamefont {Steinhardt}},\
  }\bibfield  {title} {\bibinfo {title} {Designer disordered materials with
  large, complete photonic band gaps},\ }\href@noop {} {\bibfield  {journal}
  {\bibinfo  {journal} {Proc. Natl. Acad. Sci. USA}\ }\textbf {\bibinfo
  {volume} {106}},\ \bibinfo {pages} {20658} (\bibinfo {year}
  {2009}{\natexlab{b}})}\BibitemShut {NoStop}%
\bibitem [{\citenamefont {Men}\ \emph {et~al.}(2014)\citenamefont {Men},
  \citenamefont {Lee}, \citenamefont {Freund}, \citenamefont {Peraire},\ and\
  \citenamefont {Johnson}}]{Men2014}%
  \BibitemOpen
  \bibfield  {author} {\bibinfo {author} {\bibfnamefont {H.}~\bibnamefont
  {Men}}, \bibinfo {author} {\bibfnamefont {K.~Y.~K.}\ \bibnamefont {Lee}},
  \bibinfo {author} {\bibfnamefont {R.~M.}\ \bibnamefont {Freund}}, \bibinfo
  {author} {\bibfnamefont {J.}~\bibnamefont {Peraire}},\ and\ \bibinfo {author}
  {\bibfnamefont {S.~G.}\ \bibnamefont {Johnson}},\ }\bibfield  {title}
  {\bibinfo {title} {Robust topology optimization of three-dimensional
  photonic-crystal band-gap structures},\ }\href@noop {} {\bibfield  {journal}
  {\bibinfo  {journal} {Opt. Express}\ }\textbf {\bibinfo {volume} {22}},\
  \bibinfo {pages} {22632} (\bibinfo {year} {2014})}\BibitemShut {NoStop}%
\bibitem [{\citenamefont {Cerjan}\ and\ \citenamefont
  {Fan}(2017)}]{CerjanFan2017}%
  \BibitemOpen
  \bibfield  {author} {\bibinfo {author} {\bibfnamefont {A.}~\bibnamefont
  {Cerjan}}\ and\ \bibinfo {author} {\bibfnamefont {S.}~\bibnamefont {Fan}},\
  }\bibfield  {title} {\bibinfo {title} {Complete photonic band gaps in
  supercell photonic crystals},\ }\href
  {https://doi.org/10.1103/PhysRevA.96.051802} {\bibfield  {journal} {\bibinfo
  {journal} {Phys. Rev. A}\ }\textbf {\bibinfo {volume} {96}},\ \bibinfo
  {pages} {051802(R)} (\bibinfo {year} {2017})}\BibitemShut {NoStop}%
\bibitem [{\citenamefont {Chen}\ \emph {et~al.}(2019)\citenamefont {Chen},
  \citenamefont {Morgan}, \citenamefont {Alzaidy}, \citenamefont {Huang},
  \citenamefont {Ho}, \citenamefont {Taverne}, \citenamefont {Zheng},
  \citenamefont {Ren}, \citenamefont {Feng}, \citenamefont {Zeimpekis},
  \citenamefont {Hewak},\ and\ \citenamefont {Rarity}}]{Chen2019}%
  \BibitemOpen
  \bibfield  {author} {\bibinfo {author} {\bibfnamefont {L.}~\bibnamefont
  {Chen}}, \bibinfo {author} {\bibfnamefont {K.~A.}\ \bibnamefont {Morgan}},
  \bibinfo {author} {\bibfnamefont {G.~A.}\ \bibnamefont {Alzaidy}}, \bibinfo
  {author} {\bibfnamefont {C.-C.}\ \bibnamefont {Huang}}, \bibinfo {author}
  {\bibfnamefont {Y.-L.~D.}\ \bibnamefont {Ho}}, \bibinfo {author}
  {\bibfnamefont {M.~P.~C.}\ \bibnamefont {Taverne}}, \bibinfo {author}
  {\bibfnamefont {X.}~\bibnamefont {Zheng}}, \bibinfo {author} {\bibfnamefont
  {Z.}~\bibnamefont {Ren}}, \bibinfo {author} {\bibfnamefont {Z.}~\bibnamefont
  {Feng}}, \bibinfo {author} {\bibfnamefont {I.}~\bibnamefont {Zeimpekis}},
  \bibinfo {author} {\bibfnamefont {D.~W.}\ \bibnamefont {Hewak}},\ and\
  \bibinfo {author} {\bibfnamefont {J.~G.}\ \bibnamefont {Rarity}},\ }\bibfield
   {title} {\bibinfo {title} {Observation of complete photonic bandgap in low
  refractive index contrast inverse rod-connected diamond structured
  chalcogenides},\ }\href {https://doi.org/10.1021/acsphotonics.9b00184}
  {\bibfield  {journal} {\bibinfo  {journal} {ACS Photonics}\ }\textbf
  {\bibinfo {volume} {6}},\ \bibinfo {pages} {1248} (\bibinfo {year}
  {2019})}\BibitemShut {NoStop}%
\bibitem [{\citenamefont {Cersonsky}\ \emph {et~al.}(2021)\citenamefont
  {Cersonsky}, \citenamefont {Antonaglia}, \citenamefont {Dice},\ and\
  \citenamefont {Glotzer}}]{Cersonsky2021}%
  \BibitemOpen
  \bibfield  {author} {\bibinfo {author} {\bibfnamefont {R.~K.}\ \bibnamefont
  {Cersonsky}}, \bibinfo {author} {\bibfnamefont {J.}~\bibnamefont
  {Antonaglia}}, \bibinfo {author} {\bibfnamefont {B.~D.}\ \bibnamefont
  {Dice}},\ and\ \bibinfo {author} {\bibfnamefont {S.~C.}\ \bibnamefont
  {Glotzer}},\ }\bibfield  {title} {\bibinfo {title} {The diversity of
  three-dimensional photonic crystals},\ }\href
  {https://doi.org/10.1038/s41467-021-22809-6} {\bibfield  {journal} {\bibinfo
  {journal} {Nat. Commun.}\ }\textbf {\bibinfo {volume} {12}},\ \bibinfo
  {pages} {2543} (\bibinfo {year} {2021})}\BibitemShut {NoStop}%
\bibitem [{\citenamefont {Christensen}\ \emph {et~al.}(2022)\citenamefont
  {Christensen}, \citenamefont {Po}, \citenamefont {Joannopoulos},\ and\
  \citenamefont {Solja{\v{c}}i{\'c}}}]{Christensen2022}%
  \BibitemOpen
  \bibfield  {author} {\bibinfo {author} {\bibfnamefont {T.}~\bibnamefont
  {Christensen}}, \bibinfo {author} {\bibfnamefont {H.~C.}\ \bibnamefont {Po}},
  \bibinfo {author} {\bibfnamefont {J.~D.}\ \bibnamefont {Joannopoulos}},\ and\
  \bibinfo {author} {\bibfnamefont {M.}~\bibnamefont {Solja{\v{c}}i{\'c}}},\
  }\bibfield  {title} {\bibinfo {title} {Location and topology of the
  fundamental gap in photonic crystals},\ }\href@noop {} {\bibfield  {journal}
  {\bibinfo  {journal} {Phys. Rev. X}\ }\textbf {\bibinfo {volume} {12}},\
  \bibinfo {pages} {021066} (\bibinfo {year} {2022})}\BibitemShut {NoStop}%
\bibitem [{\citenamefont {Maiwald}\ \emph {et~al.}(2022)\citenamefont
  {Maiwald}, \citenamefont {Sommer}, \citenamefont {Sidorenko}, \citenamefont
  {Yafyasov}, \citenamefont {Mustafa}, \citenamefont {Schulz}, \citenamefont
  {Rybin}, \citenamefont {Eich},\ and\ \citenamefont {Petrov}}]{Maiwald2022}%
  \BibitemOpen
  \bibfield  {author} {\bibinfo {author} {\bibfnamefont {L.}~\bibnamefont
  {Maiwald}}, \bibinfo {author} {\bibfnamefont {T.}~\bibnamefont {Sommer}},
  \bibinfo {author} {\bibfnamefont {M.~S.}\ \bibnamefont {Sidorenko}}, \bibinfo
  {author} {\bibfnamefont {R.~R.}\ \bibnamefont {Yafyasov}}, \bibinfo {author}
  {\bibfnamefont {M.~E.}\ \bibnamefont {Mustafa}}, \bibinfo {author}
  {\bibfnamefont {M.}~\bibnamefont {Schulz}}, \bibinfo {author} {\bibfnamefont
  {M.~V.}\ \bibnamefont {Rybin}}, \bibinfo {author} {\bibfnamefont
  {M.}~\bibnamefont {Eich}},\ and\ \bibinfo {author} {\bibfnamefont {A.~Y.}\
  \bibnamefont {Petrov}},\ }\bibfield  {title} {\bibinfo {title} {Control over
  light emission in low-refractive-index artificial materials inspired by
  reciprocal design},\ }\href {https://doi.org/10.1002/adom.202100785}
  {\bibfield  {journal} {\bibinfo  {journal} {Adv. Opt. Mater.}\ }\textbf
  {\bibinfo {volume} {10}},\ \bibinfo {pages} {2100785} (\bibinfo {year}
  {2022})}\BibitemShut {NoStop}%
\bibitem [{\citenamefont {Saha}\ \emph {et~al.}(2023)\citenamefont {Saha},
  \citenamefont {Mustafa}, \citenamefont {Eich},\ and\ \citenamefont
  {Petrov}}]{Saha2023}%
  \BibitemOpen
  \bibfield  {author} {\bibinfo {author} {\bibfnamefont {S.}~\bibnamefont
  {Saha}}, \bibinfo {author} {\bibfnamefont {M.~E.}\ \bibnamefont {Mustafa}},
  \bibinfo {author} {\bibfnamefont {M.}~\bibnamefont {Eich}},\ and\ \bibinfo
  {author} {\bibfnamefont {A.~Y.}\ \bibnamefont {Petrov}},\ }\bibfield  {title}
  {\bibinfo {title} {10 {dB} emission suppression in a structured low index
  medium},\ }\href {https://doi.org/10.1364/JOSAB.477576} {\bibfield  {journal}
  {\bibinfo  {journal} {J. Opt. Soc. Am. B}\ }\textbf {\bibinfo {volume}
  {40}},\ \bibinfo {pages} {900} (\bibinfo {year} {2023})}\BibitemShut
  {NoStop}%
\bibitem [{\citenamefont {Siedentop}\ \emph {et~al.}(2024)\citenamefont
  {Siedentop}, \citenamefont {Lui}, \citenamefont {Maret}, \citenamefont
  {Chaikin}, \citenamefont {Steinhardt}, \citenamefont {Torquato},
  \citenamefont {Keim},\ and\ \citenamefont {Florescu}}]{Siedentop2024}%
  \BibitemOpen
  \bibfield  {author} {\bibinfo {author} {\bibfnamefont {L.}~\bibnamefont
  {Siedentop}}, \bibinfo {author} {\bibfnamefont {G.}~\bibnamefont {Lui}},
  \bibinfo {author} {\bibfnamefont {G.}~\bibnamefont {Maret}}, \bibinfo
  {author} {\bibfnamefont {P.~M.}\ \bibnamefont {Chaikin}}, \bibinfo {author}
  {\bibfnamefont {P.~J.}\ \bibnamefont {Steinhardt}}, \bibinfo {author}
  {\bibfnamefont {S.}~\bibnamefont {Torquato}}, \bibinfo {author}
  {\bibfnamefont {P.}~\bibnamefont {Keim}},\ and\ \bibinfo {author}
  {\bibfnamefont {M.}~\bibnamefont {Florescu}},\ }\bibfield  {title} {\bibinfo
  {title} {Stealthy and hyperuniform isotropic photonic band gap structure in
  {3D}},\ }\href {https://doi.org/10.1093/pnasnexus/pgae383} {\bibfield
  {journal} {\bibinfo  {journal} {PNAS Nexus}\ }\textbf {\bibinfo {volume}
  {3}},\ \bibinfo {pages} {pgae383} (\bibinfo {year} {2024})}\BibitemShut
  {NoStop}%
\bibitem [{\citenamefont {Zhang}\ \emph {et~al.}(2024)\citenamefont {Zhang},
  \citenamefont {Min}, \citenamefont {Wang}, \citenamefont {Wang},
  \citenamefont {Li}, \citenamefont {Ha}, \citenamefont {Zhang}, \citenamefont
  {Pan}, \citenamefont {Li}, \citenamefont {Liu}, \citenamefont {Yin},
  \citenamefont {Yang}, \citenamefont {Liu}, \citenamefont {Xu}, \citenamefont
  {He}, \citenamefont {Yang},\ and\ \citenamefont {Yang}}]{Zhang2024}%
  \BibitemOpen
  \bibfield  {author} {\bibinfo {author} {\bibfnamefont {W.}~\bibnamefont
  {Zhang}}, \bibinfo {author} {\bibfnamefont {J.}~\bibnamefont {Min}}, \bibinfo
  {author} {\bibfnamefont {H.}~\bibnamefont {Wang}}, \bibinfo {author}
  {\bibfnamefont {H.}~\bibnamefont {Wang}}, \bibinfo {author} {\bibfnamefont
  {X.~L.}\ \bibnamefont {Li}}, \bibinfo {author} {\bibfnamefont {S.~T.}\
  \bibnamefont {Ha}}, \bibinfo {author} {\bibfnamefont {B.}~\bibnamefont
  {Zhang}}, \bibinfo {author} {\bibfnamefont {C.-F.}\ \bibnamefont {Pan}},
  \bibinfo {author} {\bibfnamefont {H.}~\bibnamefont {Li}}, \bibinfo {author}
  {\bibfnamefont {H.}~\bibnamefont {Liu}}, \bibinfo {author} {\bibfnamefont
  {H.}~\bibnamefont {Yin}}, \bibinfo {author} {\bibfnamefont {X.}~\bibnamefont
  {Yang}}, \bibinfo {author} {\bibfnamefont {S.}~\bibnamefont {Liu}}, \bibinfo
  {author} {\bibfnamefont {X.}~\bibnamefont {Xu}}, \bibinfo {author}
  {\bibfnamefont {C.}~\bibnamefont {He}}, \bibinfo {author} {\bibfnamefont
  {H.~Y.}\ \bibnamefont {Yang}},\ and\ \bibinfo {author} {\bibfnamefont
  {J.~K.~W.}\ \bibnamefont {Yang}},\ }\bibfield  {title} {\bibinfo {title}
  {Printing of {3D} photonic crystals in titania with complete bandgap across
  the visible spectrum},\ }\href {https://doi.org/10.1038/s41565-024-01780-5}
  {\bibfield  {journal} {\bibinfo  {journal} {Nat. Nanotechnol.}\ }\textbf
  {\bibinfo {volume} {19}},\ \bibinfo {pages} {1813} (\bibinfo {year}
  {2024})}\BibitemShut {NoStop}%
\bibitem [{\citenamefont {Casiulis}\ \emph {et~al.}(2025)\citenamefont
  {Casiulis}, \citenamefont {Shih},\ and\ \citenamefont
  {Martiniani}}]{Casiulis2025}%
  \BibitemOpen
  \bibfield  {author} {\bibinfo {author} {\bibfnamefont {M.}~\bibnamefont
  {Casiulis}}, \bibinfo {author} {\bibfnamefont {A.}~\bibnamefont {Shih}},\
  and\ \bibinfo {author} {\bibfnamefont {S.}~\bibnamefont {Martiniani}},\
  }\bibfield  {title} {\bibinfo {title} {Gyromorphs: a new class of functional
  disordered materials},\ }\href {https://doi.org/10.1103/gqrx-7mn2} {\bibfield
   {journal} {\bibinfo  {journal} {Phys. Rev. Lett.}\ }\textbf {\bibinfo
  {volume} {135}},\ \bibinfo {pages} {196101} (\bibinfo {year}
  {2025})}\BibitemShut {NoStop}%
\bibitem [{\citenamefont {Itin}(2025)}]{Itin2025}%
  \BibitemOpen
  \bibfield  {author} {\bibinfo {author} {\bibfnamefont {A.}~\bibnamefont
  {Itin}},\ }\bibfield  {title} {\bibinfo {title} {Geometric approach to design
  of photonic crystals and metamaterials via optimal toric packings},\
  }\href@noop {} {\bibfield  {journal} {\bibinfo  {journal} {Phys. Rev. B}\
  }\textbf {\bibinfo {volume} {111}},\ \bibinfo {pages} {L121404} (\bibinfo
  {year} {2025})},\ \Eprint {https://arxiv.org/abs/2410.17424} {2410.17424}
  \BibitemShut {NoStop}%
\bibitem [{\citenamefont {Vukusic}\ and\ \citenamefont
  {Sambles}(2003)}]{VukusicSambles2003}%
  \BibitemOpen
  \bibfield  {author} {\bibinfo {author} {\bibfnamefont {P.}~\bibnamefont
  {Vukusic}}\ and\ \bibinfo {author} {\bibfnamefont {J.~R.}\ \bibnamefont
  {Sambles}},\ }\bibfield  {title} {\bibinfo {title} {Photonic structures in
  biology},\ }\href {https://doi.org/10.1038/nature01941} {\bibfield  {journal}
  {\bibinfo  {journal} {Nature}\ }\textbf {\bibinfo {volume} {424}},\ \bibinfo
  {pages} {852} (\bibinfo {year} {2003})}\BibitemShut {NoStop}%
\bibitem [{\citenamefont {Galusha}\ \emph {et~al.}(2008)\citenamefont
  {Galusha}, \citenamefont {Richey}, \citenamefont {Gardner}, \citenamefont
  {Cha},\ and\ \citenamefont {Bartl}}]{Galusha2008}%
  \BibitemOpen
  \bibfield  {author} {\bibinfo {author} {\bibfnamefont {J.~W.}\ \bibnamefont
  {Galusha}}, \bibinfo {author} {\bibfnamefont {L.~R.}\ \bibnamefont {Richey}},
  \bibinfo {author} {\bibfnamefont {J.~S.}\ \bibnamefont {Gardner}}, \bibinfo
  {author} {\bibfnamefont {J.~N.}\ \bibnamefont {Cha}},\ and\ \bibinfo {author}
  {\bibfnamefont {M.~H.}\ \bibnamefont {Bartl}},\ }\bibfield  {title} {\bibinfo
  {title} {Discovery of a diamond-based photonic crystal structure in beetle
  scales},\ }\href {https://doi.org/10.1103/PhysRevE.77.050904} {\bibfield
  {journal} {\bibinfo  {journal} {Phys. Rev. E}\ }\textbf {\bibinfo {volume}
  {77}},\ \bibinfo {pages} {050904} (\bibinfo {year} {2008})}\BibitemShut
  {NoStop}%
\bibitem [{\citenamefont {{Lord Rayleigh}}(1887)}]{Rayleigh1887}%
  \BibitemOpen
  \bibfield  {author} {\bibinfo {author} {\bibnamefont {{Lord Rayleigh}}},\
  }\bibfield  {title} {\bibinfo {title} {On the maintenance of vibrations by
  forces of double frequency, and on the propagation of waves through a medium
  endowed with a periodic structure},\ }\href
  {https://doi.org/10.1080/14786448708628074} {\bibfield  {journal} {\bibinfo
  {journal} {Philos. Mag.}\ }\textbf {\bibinfo {volume} {24}},\ \bibinfo
  {pages} {145} (\bibinfo {year} {1887})}\BibitemShut {NoStop}%
\bibitem [{\citenamefont {Bykov}(1972)}]{Bykov1972}%
  \BibitemOpen
  \bibfield  {author} {\bibinfo {author} {\bibfnamefont {V.~P.}\ \bibnamefont
  {Bykov}},\ }\bibfield  {title} {\bibinfo {title} {Spontaneous emission in a
  periodic structure},\ }\href@noop {} {\bibfield  {journal} {\bibinfo
  {journal} {Sov. Phys. JETP}\ }\textbf {\bibinfo {volume} {35}},\ \bibinfo
  {pages} {269} (\bibinfo {year} {1972})}\BibitemShut {NoStop}%
\bibitem [{\citenamefont {Gissibl}\ \emph {et~al.}(2017)\citenamefont
  {Gissibl}, \citenamefont {Wagner}, \citenamefont {Sykora}, \citenamefont
  {Schmid},\ and\ \citenamefont {Giessen}}]{Gissibl2017}%
  \BibitemOpen
  \bibfield  {author} {\bibinfo {author} {\bibfnamefont {T.}~\bibnamefont
  {Gissibl}}, \bibinfo {author} {\bibfnamefont {S.}~\bibnamefont {Wagner}},
  \bibinfo {author} {\bibfnamefont {J.}~\bibnamefont {Sykora}}, \bibinfo
  {author} {\bibfnamefont {M.}~\bibnamefont {Schmid}},\ and\ \bibinfo {author}
  {\bibfnamefont {H.}~\bibnamefont {Giessen}},\ }\bibfield  {title} {\bibinfo
  {title} {Refractive index measurements of photo-resists for three-dimensional
  direct laser writing},\ }\href {https://doi.org/10.1364/OME.7.002293}
  {\bibfield  {journal} {\bibinfo  {journal} {Opt. Mater. Express}\ }\textbf
  {\bibinfo {volume} {7}},\ \bibinfo {pages} {2293} (\bibinfo {year}
  {2017})}\BibitemShut {NoStop}%
\bibitem [{\citenamefont {Jones}(1934)}]{Jones1934}%
  \BibitemOpen
  \bibfield  {author} {\bibinfo {author} {\bibfnamefont {H.}~\bibnamefont
  {Jones}},\ }\bibfield  {title} {\bibinfo {title} {The theory of alloys in the
  $\gamma$-phase},\ }\href {https://doi.org/10.1098/rspa.1934.0044} {\bibfield
  {journal} {\bibinfo  {journal} {Proc. R. Soc. Lond. A}\ }\textbf {\bibinfo
  {volume} {144}},\ \bibinfo {pages} {225} (\bibinfo {year}
  {1934})}\BibitemShut {NoStop}%
\bibitem [{SM()}]{SM}%
  \BibitemOpen
  \href@noop {} {}\bibinfo {note} {See the Supplemental Material after the
  references (Secs.~S1--S13) for the numerical effort (S1), the solver and the
  optimizer (S2, S3), the band-edge search, the numerical validation and the
  covering certificate (S4, S5), the verification data of every design (S6),
  the empty-lattice bound (S7), the reciprocal-space analysis and the counting
  estimate (S8, S9), data and code (S10), decay inside the gap, connectivity
  and a printed sample (S11), the polyhedral starts (S12), and the
  two-dimensional designs (S13)}\BibitemShut {NoStop}%
\bibitem [{\citenamefont {Gravin}\ \emph {et~al.}(2012)\citenamefont {Gravin},
  \citenamefont {Robins},\ and\ \citenamefont
  {Shiryaev}}]{GravinRobinsShiryaev2012}%
  \BibitemOpen
  \bibfield  {author} {\bibinfo {author} {\bibfnamefont {N.}~\bibnamefont
  {Gravin}}, \bibinfo {author} {\bibfnamefont {S.}~\bibnamefont {Robins}},\
  and\ \bibinfo {author} {\bibfnamefont {D.}~\bibnamefont {Shiryaev}},\
  }\bibfield  {title} {\bibinfo {title} {Translational tilings by a polytope,
  with multiplicity},\ }\href {https://doi.org/10.1007/s00493-012-2860-3}
  {\bibfield  {journal} {\bibinfo  {journal} {Combinatorica}\ }\textbf
  {\bibinfo {volume} {32}},\ \bibinfo {pages} {629} (\bibinfo {year}
  {2012})}\BibitemShut {NoStop}%
\bibitem [{\citenamefont {Johnson}\ and\ \citenamefont
  {Joannopoulos}(2001)}]{JohnsonJoannopoulos2001}%
  \BibitemOpen
  \bibfield  {author} {\bibinfo {author} {\bibfnamefont {S.~G.}\ \bibnamefont
  {Johnson}}\ and\ \bibinfo {author} {\bibfnamefont {J.~D.}\ \bibnamefont
  {Joannopoulos}},\ }\bibfield  {title} {\bibinfo {title} {Block-iterative
  frequency-domain methods for {Maxwell's} equations in a planewave basis},\
  }\href@noop {} {\bibfield  {journal} {\bibinfo  {journal} {Opt. Express}\
  }\textbf {\bibinfo {volume} {8}},\ \bibinfo {pages} {173} (\bibinfo {year}
  {2001})}\BibitemShut {NoStop}%
\bibitem [{\citenamefont {Musin}\ and\ \citenamefont
  {Tarasov}(2015)}]{MusinTarasov2015}%
  \BibitemOpen
  \bibfield  {author} {\bibinfo {author} {\bibfnamefont {O.~R.}\ \bibnamefont
  {Musin}}\ and\ \bibinfo {author} {\bibfnamefont {A.~S.}\ \bibnamefont
  {Tarasov}},\ }\bibfield  {title} {\bibinfo {title} {The {T}ammes problem for
  {$N = 14$}},\ }\href@noop {} {\bibfield  {journal} {\bibinfo  {journal} {Exp.
  Math.}\ }\textbf {\bibinfo {volume} {24}},\ \bibinfo {pages} {460} (\bibinfo
  {year} {2015})}\BibitemShut {NoStop}%
\bibitem [{\citenamefont {Conway}\ and\ \citenamefont
  {Sloane}(1993)}]{ConwaySloane1993}%
  \BibitemOpen
  \bibfield  {author} {\bibinfo {author} {\bibfnamefont {J.~H.}\ \bibnamefont
  {Conway}}\ and\ \bibinfo {author} {\bibfnamefont {N.~J.~A.}\ \bibnamefont
  {Sloane}},\ }\href@noop {} {\emph {\bibinfo {title} {Sphere Packings,
  Lattices and Groups}}},\ \bibinfo {edition} {2nd}\ ed.\ (\bibinfo
  {publisher} {Springer},\ \bibinfo {address} {New York},\ \bibinfo {year}
  {1993})\BibitemShut {NoStop}%
\bibitem [{\citenamefont {Conway}\ \emph {et~al.}(1996)\citenamefont {Conway},
  \citenamefont {Hardin},\ and\ \citenamefont
  {Sloane}}]{ConwayHardinSloane1996}%
  \BibitemOpen
  \bibfield  {author} {\bibinfo {author} {\bibfnamefont {J.~H.}\ \bibnamefont
  {Conway}}, \bibinfo {author} {\bibfnamefont {R.~H.}\ \bibnamefont {Hardin}},\
  and\ \bibinfo {author} {\bibfnamefont {N.~J.~A.}\ \bibnamefont {Sloane}},\
  }\bibfield  {title} {\bibinfo {title} {Packing lines, planes, etc.: packings
  in {G}rassmannian spaces},\ }\href@noop {} {\bibfield  {journal} {\bibinfo
  {journal} {Exp. Math.}\ }\textbf {\bibinfo {volume} {5}},\ \bibinfo {pages}
  {139} (\bibinfo {year} {1996})}\BibitemShut {NoStop}%
\bibitem [{\citenamefont {Hardin}\ \emph {et~al.}(1994)\citenamefont {Hardin},
  \citenamefont {Sloane},\ and\ \citenamefont {Smith}}]{HardinSloaneSmith}%
  \BibitemOpen
  \bibfield  {author} {\bibinfo {author} {\bibfnamefont {R.~H.}\ \bibnamefont
  {Hardin}}, \bibinfo {author} {\bibfnamefont {N.~J.~A.}\ \bibnamefont
  {Sloane}},\ and\ \bibinfo {author} {\bibfnamefont {W.~D.}\ \bibnamefont
  {Smith}},\ }\href@noop {} {\bibinfo {title} {Tables of spherical codes with
  icosahedral symmetry}},\ \bibinfo {howpublished}
  {\url{http://neilsloane.com/icosahedral.codes}; now maintained by H. Cohn,
  \url{https://cohn.mit.edu/sloane}} (\bibinfo {year} {1994})\BibitemShut
  {NoStop}%
\bibitem [{\citenamefont {Weaire}(1999)}]{Weaire1999}%
  \BibitemOpen
  \bibfield  {author} {\bibinfo {author} {\bibfnamefont {D.}~\bibnamefont
  {Weaire}},\ }\bibfield  {title} {\bibinfo {title} {A short history of packing
  problems},\ }\href@noop {} {\bibfield  {journal} {\bibinfo  {journal}
  {Forma}\ }\textbf {\bibinfo {volume} {14}},\ \bibinfo {pages} {279} (\bibinfo
  {year} {1999})}\BibitemShut {NoStop}%
\bibitem [{\citenamefont {Aste}\ and\ \citenamefont
  {Weaire}(2000)}]{AsteWeaire2000}%
  \BibitemOpen
  \bibfield  {author} {\bibinfo {author} {\bibfnamefont {T.}~\bibnamefont
  {Aste}}\ and\ \bibinfo {author} {\bibfnamefont {D.}~\bibnamefont {Weaire}},\
  }\href@noop {} {\emph {\bibinfo {title} {The Pursuit of Perfect Packing}}}\
  (\bibinfo  {publisher} {Institute of Physics Publishing},\ \bibinfo {address}
  {Bristol},\ \bibinfo {year} {2000})\BibitemShut {NoStop}%
\bibitem [{\citenamefont {Musin}\ and\ \citenamefont
  {Nikitenko}(2016)}]{MusinNikitenko2016}%
  \BibitemOpen
  \bibfield  {author} {\bibinfo {author} {\bibfnamefont {O.~R.}\ \bibnamefont
  {Musin}}\ and\ \bibinfo {author} {\bibfnamefont {A.~V.}\ \bibnamefont
  {Nikitenko}},\ }\bibfield  {title} {\bibinfo {title} {Optimal packings of
  congruent circles on a square flat torus},\ }\href@noop {} {\bibfield
  {journal} {\bibinfo  {journal} {Discrete Comput. Geom.}\ }\textbf {\bibinfo
  {volume} {55}},\ \bibinfo {pages} {1} (\bibinfo {year} {2016})}\BibitemShut
  {NoStop}%
\bibitem [{\citenamefont {Connelly}\ and\ \citenamefont
  {Dickinson}(2014)}]{ConnellyDickinson2014}%
  \BibitemOpen
  \bibfield  {author} {\bibinfo {author} {\bibfnamefont {R.}~\bibnamefont
  {Connelly}}\ and\ \bibinfo {author} {\bibfnamefont {W.}~\bibnamefont
  {Dickinson}},\ }\bibfield  {title} {\bibinfo {title} {Periodic planar disc
  packings},\ }\href@noop {} {\bibfield  {journal} {\bibinfo  {journal}
  {Philos. Trans. R. Soc. A}\ }\textbf {\bibinfo {volume} {372}},\ \bibinfo
  {pages} {20120039} (\bibinfo {year} {2014})}\BibitemShut {NoStop}%
\bibitem [{\citenamefont {P{\'e}rez-Escribano}\ and\ \citenamefont
  {M{\'a}rquez-Segura}(2021)}]{PerezEscribano2021}%
  \BibitemOpen
  \bibfield  {author} {\bibinfo {author} {\bibfnamefont {M.}~\bibnamefont
  {P{\'e}rez-Escribano}}\ and\ \bibinfo {author} {\bibfnamefont
  {E.}~\bibnamefont {M{\'a}rquez-Segura}},\ }\bibfield  {title} {\bibinfo
  {title} {Parameters characterization of dielectric materials samples in
  microwave and millimeter-wave bands},\ }\href
  {https://doi.org/10.1109/TMTT.2020.3045211} {\bibfield  {journal} {\bibinfo
  {journal} {IEEE Trans. Microwave Theory Tech.}\ }\textbf {\bibinfo {volume}
  {69}},\ \bibinfo {pages} {1723} (\bibinfo {year} {2021})}\BibitemShut
  {NoStop}%
\bibitem [{\citenamefont {Deubel}\ \emph {et~al.}(2004)\citenamefont {Deubel},
  \citenamefont {von Freymann}, \citenamefont {Wegener}, \citenamefont
  {Pereira}, \citenamefont {Busch},\ and\ \citenamefont
  {Soukoulis}}]{Deubel2004}%
  \BibitemOpen
  \bibfield  {author} {\bibinfo {author} {\bibfnamefont {M.}~\bibnamefont
  {Deubel}}, \bibinfo {author} {\bibfnamefont {G.}~\bibnamefont {von
  Freymann}}, \bibinfo {author} {\bibfnamefont {M.}~\bibnamefont {Wegener}},
  \bibinfo {author} {\bibfnamefont {S.}~\bibnamefont {Pereira}}, \bibinfo
  {author} {\bibfnamefont {K.}~\bibnamefont {Busch}},\ and\ \bibinfo {author}
  {\bibfnamefont {C.~M.}\ \bibnamefont {Soukoulis}},\ }\bibfield  {title}
  {\bibinfo {title} {Direct laser writing of three-dimensional photonic-crystal
  templates for telecommunications},\ }\href@noop {} {\bibfield  {journal}
  {\bibinfo  {journal} {Nat. Mater.}\ }\textbf {\bibinfo {volume} {3}},\
  \bibinfo {pages} {444} (\bibinfo {year} {2004})}\BibitemShut {NoStop}%
\bibitem [{\citenamefont {Hu}\ \emph {et~al.}(2026)\citenamefont {Hu},
  \citenamefont {Li}, \citenamefont {Gao}, \citenamefont {Chen}, \citenamefont
  {Zhang}, \citenamefont {Li}, \citenamefont {Mao},\ and\ \citenamefont
  {Xiong}}]{Hu2026}%
  \BibitemOpen
  \bibfield  {author} {\bibinfo {author} {\bibfnamefont {H.}~\bibnamefont
  {Hu}}, \bibinfo {author} {\bibfnamefont {D.}~\bibnamefont {Li}}, \bibinfo
  {author} {\bibfnamefont {H.}~\bibnamefont {Gao}}, \bibinfo {author}
  {\bibfnamefont {F.}~\bibnamefont {Chen}}, \bibinfo {author} {\bibfnamefont
  {M.}~\bibnamefont {Zhang}}, \bibinfo {author} {\bibfnamefont
  {Z.}~\bibnamefont {Li}}, \bibinfo {author} {\bibfnamefont {A.}~\bibnamefont
  {Mao}},\ and\ \bibinfo {author} {\bibfnamefont {W.}~\bibnamefont {Xiong}},\
  }\bibfield  {title} {\bibinfo {title} {Controllable fabrication of
  visible-spectrum {3D} photonic crystals},\ }\href@noop {} {\bibfield
  {journal} {\bibinfo  {journal} {Sci. Adv.}\ }\textbf {\bibinfo {volume}
  {12}},\ \bibinfo {pages} {eaej0287} (\bibinfo {year} {2026})}\BibitemShut
  {NoStop}%
\bibitem [{\citenamefont {Nikulin}\ \emph {et~al.}(2022)\citenamefont
  {Nikulin}, \citenamefont {Zisman}, \citenamefont {Eich}, \citenamefont
  {Petrov},\ and\ \citenamefont {Itin}}]{Nikulin2022}%
  \BibitemOpen
  \bibfield  {author} {\bibinfo {author} {\bibfnamefont {A.}~\bibnamefont
  {Nikulin}}, \bibinfo {author} {\bibfnamefont {I.}~\bibnamefont {Zisman}},
  \bibinfo {author} {\bibfnamefont {M.}~\bibnamefont {Eich}}, \bibinfo {author}
  {\bibfnamefont {A.~Y.}\ \bibnamefont {Petrov}},\ and\ \bibinfo {author}
  {\bibfnamefont {A.}~\bibnamefont {Itin}},\ }\bibfield  {title} {\bibinfo
  {title} {Machine learning models for photonic crystals band diagram
  prediction and gap optimisation},\ }\href
  {https://doi.org/10.1016/j.photonics.2022.101076} {\bibfield  {journal}
  {\bibinfo  {journal} {Photonics Nanostruct. Fundam. Appl.}\ }\textbf
  {\bibinfo {volume} {52}},\ \bibinfo {pages} {101076} (\bibinfo {year}
  {2022})}\BibitemShut {NoStop}%
\bibitem [{\citenamefont {Schurov}\ \emph {et~al.}(2024)\citenamefont
  {Schurov}, \citenamefont {Zhdanov}, \citenamefont {Alforov}, \citenamefont
  {Katsnelson}, \citenamefont {Bagrov},\ and\ \citenamefont
  {Itin}}]{Schurov2024}%
  \BibitemOpen
  \bibfield  {author} {\bibinfo {author} {\bibfnamefont {I.}~\bibnamefont
  {Schurov}}, \bibinfo {author} {\bibfnamefont {M.}~\bibnamefont {Zhdanov}},
  \bibinfo {author} {\bibfnamefont {D.}~\bibnamefont {Alforov}}, \bibinfo
  {author} {\bibfnamefont {M.}~\bibnamefont {Katsnelson}}, \bibinfo {author}
  {\bibfnamefont {A.}~\bibnamefont {Bagrov}},\ and\ \bibinfo {author}
  {\bibfnamefont {A.}~\bibnamefont {Itin}},\ }\href@noop {} {\bibinfo {title}
  {Invariant multiscale neural networks for data-scarce scientific
  applications}},\ \bibinfo {howpublished} {arXiv:2406.08318} (\bibinfo {year}
  {2024})\BibitemShut {NoStop}%
\bibitem [{\citenamefont {Hoang}\ \emph {et~al.}(2009)\citenamefont {Hoang},
  \citenamefont {Plum},\ and\ \citenamefont {Wieners}}]{HoangPlumWieners2009}%
  \BibitemOpen
  \bibfield  {author} {\bibinfo {author} {\bibfnamefont {V.}~\bibnamefont
  {Hoang}}, \bibinfo {author} {\bibfnamefont {M.}~\bibnamefont {Plum}},\ and\
  \bibinfo {author} {\bibfnamefont {C.}~\bibnamefont {Wieners}},\ }\bibfield
  {title} {\bibinfo {title} {A computer-assisted proof for photonic band
  gaps},\ }\href {https://doi.org/10.1007/s00033-008-8021-2} {\bibfield
  {journal} {\bibinfo  {journal} {Z. Angew. Math. Phys.}\ }\textbf {\bibinfo
  {volume} {60}},\ \bibinfo {pages} {1035} (\bibinfo {year}
  {2009})}\BibitemShut {NoStop}%
\bibitem [{\citenamefont {Rechtsman}\ and\ \citenamefont
  {Torquato}(2009)}]{RechtsmanTorquato2009}%
  \BibitemOpen
  \bibfield  {author} {\bibinfo {author} {\bibfnamefont {M.~C.}\ \bibnamefont
  {Rechtsman}}\ and\ \bibinfo {author} {\bibfnamefont {S.}~\bibnamefont
  {Torquato}},\ }\bibfield  {title} {\bibinfo {title} {Method for obtaining
  upper bounds on photonic band gaps},\ }\href@noop {} {\bibfield  {journal}
  {\bibinfo  {journal} {Phys. Rev. B}\ }\textbf {\bibinfo {volume} {80}},\
  \bibinfo {pages} {155126} (\bibinfo {year} {2009})}\BibitemShut {NoStop}%
\end{thebibliography}%
\clearpage
\onecolumngrid
\begin{center}{\large\bfseries Supplemental Material}\end{center}
\bigskip
\setcounter{figure}{0}\setcounter{table}{0}\setcounter{equation}{0}
\renewcommand{\thefigure}{S\arabic{figure}}\renewcommand{\theequation}{S\arabic{equation}}

\renewcommand{\thefigure}{S\arabic{figure}}
\section*{S1. Numerical effort}

All calculations ran on the TCM cluster of Radboud University, about 50 000 core-hours in total by Slurm accounting (Table S1): about 22 300 for plane-wave design, 22 700 for MPB and 1900 for the certificates of the 3D designs, and 2600 for the 2D designs of S13. The optimizer\textquotesingle s path is kept: 84 092 binary designs with bands on a coarse grid and 237 786 iterates (1.1 GB), and, for the later campaigns, also every scan, edge search and contrast search (S10).

\begin{table}[!htbp]
\renewcommand{\thetable}{S1}
\caption{Slurm core-hours by stage.}
\centering\small
\begin{tabularx}{\linewidth}{>{\raggedright\arraybackslash\hsize=1.00\hsize}Xcc}
\hline\hline
Stage & Campaigns & Core-hours \\
\hline
First designs on sparse k-sets, superseded (S4) & 6 & 6530 \\
Dense re-check of those designs (S4) & 1 & 390 \\
Design with dense band edges: Riesz, shell, random and cylindrical-wave starts; two push rounds & 4 & 3500 \\
Larger cells (n = 146--226); two push rounds at bcc 208 & 4 & 3810 \\
Polyhedral starts, screens of bare sets, push rounds (S12) & 9 & 6110 \\
Restricted-support series (S9) & 1 & 1920 \\
First MPB checks, contrast ladders, resolution study & 9 & 1890 \\
MPB claims (S5, S6) & 16 & 11 460 \\
MPB opening contrasts (S6) & 1 & 600 \\
MPB global edge searches and probes (S5) & 7 & 4870 \\
MPB band structures and densities of states & 4 & 3910 \\
Covering certificates of the plane-wave models (S5) & 4 & 1900 \\
All 3D campaigns & 66 & 46 880 \\
2D campaigns (S13) & 3 & 2570 \\
All campaigns & 69 & 49 450 \\
\hline\hline
\end{tabularx}
\end{table}

Core-hours are the CPU time Slurm allocated to each campaign\textquotesingle s jobs (CPUs \ensuremath{\times} wall time), rounded to 10. They include idle cores of the one-node jobs and stopped or unfinished cases, so the computing actually used is lower.

Runs still in progress at submission and the earliest tests bring the total to about 50 000 core-hours.

\textbf{Hardware and software.} Nodes with 20--40 cores and 22--204 GB of memory; MPB 1.12.0 through the Meep 1.35.0 Python interface; our own plane-wave code (numpy, scipy). Each campaign runs as one-node jobs (at most one per node and campaign) that work through a shared case table.

\section*{S2. Structure and plane-wave solver}

Designs are thresholded Fourier series on unit-volume primitive cells, solved for the full vector field with a truncated plane-wave basis at 2\(k_{n}\); the truncation was chosen by a cutoff study.

\textbf{Cells and units.} The sc, fcc and bcc primitive cells are scaled to unit volume. Reciprocal vectors G = m\ensuremath{\cdot}B (m integer) carry no factor 2\ensuremath{\pi}, so plane waves are exp(2\ensuremath{\pi}i G\ensuremath{\cdot}r), and k is given in fractional reciprocal coordinates. Band n of the empty lattice lies at \textbar k\textbar{} = \(k_{n}\) = (3n/8\ensuremath{\pi})\(^{1/3}\); the Bragg sphere has radius R = 2\(k_{n}\).

\textbf{Symmetry.} The lattice is the only symmetry imposed. The Fourier support consists of reciprocal vectors of the chosen cell, so every design keeps its translations exactly, but the coefficients are free, and neither the start\textquotesingle s point group nor inversion survives the optimization. For each of the 48 cubic operations R we computed the best overlap of \ensuremath{\theta}(r) with \ensuremath{\theta}(Rr + t) over all grid translations t, scaled so that an exact symmetry gives 1. The best non-identity operation reaches 0.42 in the design at index 1.547 (inversion: 0.21), 0.30 in the IDIC design at 1.594, 0.24 on fcc 118, 0.36 on sc 88 and 0.17 on bcc 190, against medians of 0.11--0.18 over the 47 non-identity operations. The designs are thus triclinic structures (space group P1) on a cubic lattice of translations; a label such as bcc 136 names that lattice and the band. A structure periodic on bcc is also periodic on the simple-cubic lattice of its conventional cube, which holds two primitive cells, so band 136 of bcc is band 272 of that description; the bcc label gives the smallest cell. Without point symmetry the irreducible zone is half the Brillouin zone, from time reversal (k \ensuremath{\rightarrow} \ensuremath{-}k) alone, which is why the band edges are searched over the whole zone (S4) and why high-symmetry paths, as in Fig. S1, only illustrate the bands.

\textbf{Level set.} With s the fractional position in the cell:

\begin{equation}
g(s) = \sum_{j=1}^{P} \mathrm{Re}\!\left[c_j\, e^{2\pi i\, m_j\cdot s}\right], \qquad \rho(s) = \sigma\!\left(\frac{g(s)/g_{\mathrm{rms}} - t}{\tau}\right), \qquad \varepsilon = 1 + (c-1)\,\rho
\end{equation}

The support holds one vector of each pair \ensuremath{\pm}m with 0.15 \ensuremath{\times} 2\(k_{n}\) \ensuremath{\leq} \textbar G\textbar{} \ensuremath{\leq} 1.06 \ensuremath{\times} 2\(k_{n}\), 313 pairs at bcc 136, 268 at fcc 118, 218 at sc 88 and 410 at fcc 168. The variables are Re \(c_{j}\), Im \(c_{j}\) and the threshold t. The binary structure, the one verified, is \ensuremath{\theta} = {[}g/\(g_{\mathrm{rms}}\) \textgreater{} t{]}; \ensuremath{\sigma} is the logistic function and \ensuremath{\tau} is annealed to 0.02. Fourier coefficients come from a 48\(^{3}\) voxel grid, with the sinc factor of voxel averaging.

\textbf{Solver.} We use the H form in a transverse basis: two unit vectors \(e_{\mathrm{ip}}\) \ensuremath{\perp} k + \(G_{i}\) per wave, with all waves \textbar k + G\textbar{} \ensuremath{\leq} (1 + w) \(k_{n}\). With w = 1 (cutoff 2\(k_{n}\)) this is about 4n waves and 8n basis states, about 1090 at n = 136. With Ho\textquotesingle s inverse rule \cite{HoChanSoukoulis1990}:

\begin{equation}
E_{ij} = \delta_{ij} + (c-1)\,\hat\theta(m_i - m_j), \qquad \eta = E^{-1}, \qquad H_{ip,jq} = |k+G_i|\,|k+G_j|\;\eta_{ij}\;(e_{ip}\cdot e_{jq})
\end{equation}

The eigenvalues are (\ensuremath{\omega}a/2\ensuremath{\pi}c)\(^{2}\) with a the unit length. A k-point costs about 0.3 s at fcc 118 and 1.7 s at sc 216.

\textbf{Cutoff.} At the contrasts of interest, waves up to about \ensuremath{\surd}c \(k_{n}\) \ensuremath{\approx} 1.65 \(k_{n}\) come down to the gap frequency, so a 1.4 \(k_{n}\) cutoff, used in the first campaign, overstates the gap:

\begin{table}[!htbp]
\renewcommand{\thetable}{S2a}
\caption{Plane-wave gap of two designs against the cutoff of the plane-wave basis.}
\centering\small
\begin{tabular}{lccc}
\hline\hline
Design, contrast & Gap at cutoff 1.4 \(k_{n}\) & 1.7 \(k_{n}\) & 2.0 \(k_{n}\) \\
\hline
bcc 136, c = 2.8 & +0.46 \% & +0.32 \% & +0.27 \% \\
fcc 118, c = 3.2 & +0.70 \% & +0.43 \% & +0.38 \% \\
\hline\hline
\end{tabular}
\end{table}

All later designs are optimized at 2.0 \(k_{n}\) and checked again at 2.3 \(k_{n}\). The reported plane-wave contrast is the worse of the two (\(c_{\mathrm{robust}}\)).

\textbf{Agreement with MPB.} The two solvers converge toward each other. Same binary structure (bcc 136, c = 2.7343), same 12 edge k-points of its claim:

\begin{table}[!htbp]
\renewcommand{\thetable}{S2b}
\caption{Band edges of one binary structure (bcc 136, c = 2.7343) on the 12 edge k-points of its claim, in the plane-wave models and in MPB.}
\centering\small
\begin{tabularx}{\linewidth}{>{\raggedright\arraybackslash\hsize=1.12\hsize}X>{\raggedright\arraybackslash\hsize=0.97\hsize}X>{\raggedright\arraybackslash\hsize=0.97\hsize}X>{\raggedright\arraybackslash\hsize=0.95\hsize}X}
\hline\hline
Model & Max band n & Min band n + 1 & Gap \\
\hline
MPB, res 48 / 64 / 80 & 2.00249 / 2.00140 / 2.00084 & 2.01150 / 2.01070 / 2.01037 & 0.449 / 0.464 / 0.475 \% \\
MPB, 1/res\(^{2}\) extrapolation & \textasciitilde{} & \textasciitilde{} & 0.488 \% \\
Plane waves, 48\(^{3}\) voxels, cutoff 2.0 / 2.3 \(k_{n}\) (design model) & 1.97760 / 1.97945 & 1.98296 / 1.98665 & 0.271 / 0.363 \% \\
Plane waves, 192\(^{3}\) voxels, cutoff 2.0 / 2.3 \(k_{n}\) & 1.97729 / 1.97920 & 1.98353 / 1.98743 & 0.315 / 0.415 \% \\
Plane waves, 96\(^{3}\) voxels, cutoff 2.6 / 3.0 \(k_{n}\) & 1.98220 / 1.98200 & 1.98975 / 1.99072 & 0.380 / 0.439 \% \\
\hline\hline
\end{tabularx}
\end{table}

Plane-wave frequencies lie 1.1 \% below MPB\textquotesingle s at every point, but bands n and n + 1 shift together to within 0.03 \%. The gap differs through truncation and voxelization only: the design model understates it by about 0.2 points, 192\(^{3}\) voxels at 2.3 \(k_{n}\) by 0.06, and 3.0 \(k_{n}\) by 0.05 against MPB\textquotesingle s resolution limit. The design model is conservative, and every claim rests on MPB (S5). Two other rules were tested on this structure and rejected. With MPB\textquotesingle s smoothed inverse-dielectric tensor (Laurent rule) the frequencies come out 4--7 \% too high at 2.0--2.6 \(k_{n}\) and no gap appears. The normal-vector factorization gives 0.39 \% at the 12 edge points at 2.0 \(k_{n}\) but agrees with Ho\textquotesingle s rule once the plane-wave set is fixed (0.30 against 0.31 \% over the scan points). Ho\textquotesingle s rule is therefore kept for design and for the certificate.

\section*{S3. Optimizer}

The optimizer lowers the contrast round by round while L-BFGS keeps a gap open on a growing set of k-points, with exact gradients in all coefficients.

\textbf{Objective.} On a working k-set K the optimizer minimizes a smooth gap measure (\ensuremath{\beta} = 400):

\begin{equation}
J = \frac{1}{\beta}\log\sum_{k\in K} e^{\beta \log\omega_n(k)} \;+\; \frac{1}{\beta}\log\sum_{k\in K} e^{-\beta \log\omega_{n+1}(k)}
\end{equation}

J \textless{} 0 means a gap on K. Hellmann--Feynman gives d\ensuremath{\lambda} = v\dag dH v. With d\ensuremath{\eta} = \ensuremath{-}\ensuremath{\eta} dE \ensuremath{\eta}, adjoint FFTs carry the derivative through \(\hat{\theta}\), the sigmoid, the RMS normalization and the synthesis, so the gradient in all 2P + 1 variables costs about one extra solve.

\textbf{Rounds.} Each round has four steps:

\begin{enumerate}
\def\labelenumi{\arabic{enumi}.}
\tightlist
\item
  binarize the design and find \(c_{\mathrm{min}}\), the lowest contrast with a 0.3 \% gap on K;
\item
  run L-BFGS (40 iterations) on the smooth design at c0 = 1 + (\(c_{\mathrm{cur}}\) \ensuremath{-} 1) \ensuremath{\times} 0.96, with \ensuremath{\tau} = max(0.02, 0.05 \ensuremath{\times} 0.6\^{}round);
\item
  scan the binary design on kgrid(8) \ensuremath{\cup} kgrid(12) \ensuremath{\cup} 200 random k and refine 4 + 4 band edges (S4);
\item
  add the refined edges and the 4 + 4 worst raw points to K.
\end{enumerate}

The initial K is kgrid(4). v2 designs ran 14 rounds from c = 4.0. Starts that never open a gap would otherwise stall, so the larger cells (r6b) raise c \ensuremath{-} 1 by 1.25 after every round without a gap, starting from c = 4.5.

\begin{table}[!htbp]
\renewcommand{\thetable}{S3a}
\caption{Start constructions.}
\centering\small
\begin{tabularx}{\linewidth}{>{\raggedright\arraybackslash\hsize=0.36\hsize}X>{\raggedright\arraybackslash\hsize=1.64\hsize}X}
\hline\hline
Start & Construction \\
\hline
random & Gaussian ring of random coefficients around 2\(k_{n}\) \\
shell & lattice vectors with 0.88 \ensuremath{\leq}\\ & G\\ & /2\(k_{n}\) \ensuremath{\leq} 1.04, equal amplitudes, phases 0 or \ensuremath{\pi} at random \\
Riesz & the same vectors, amplitudes \ensuremath{\surd}\(q_{j}\), with \(q_{j}\) the log-Riesz equilibrium charges of the directions \(G_{j}\)/\\ & \(G_{j}\)\\ & and their antipodes (NNLS, S8) \\
cylindrical waves & \(\ensuremath{\Sigma}_{a}\) J\(_{0}\)(K\(\ensuremath{\rho}_{a}\)) about axes \(u_{a}\), K = 0.97 \ensuremath{\times} 2\(k_{n}\); weight on great circles of angular half-width 0.12 rad; 3, 4, 6 or 10 axes (cube axes, cube diagonals, icosahedral five- and three-fold axes) \\
warm / push & a finished design, optionally with Gaussian noise of 10--30 \% of the coefficient RMS, continued with c \ensuremath{-} 1 \ensuremath{\times} 0.97 per round \\
polyhedral (S12) & the lattice vectors of one or more orbits near the vertices of a polyhedron on the Bragg sphere, equal amplitudes with the orbit\textquotesingle s signs (sym) or a random sign per pair (rand), optionally over a Riesz-weighted shell at 0.3 of the peak amplitude (bg); 5 \% complex noise near the shell \\
\hline\hline
\end{tabularx}
\end{table}

\textbf{Pruning.} To test whether weak components matter, the K = 60, 40 or 25 largest pairs of a finished design are kept and re-optimized from 1.1 \(c_{03}\); the others are fixed at zero (L-BFGS-B bounds). No pruned design keeps a gap (S8).

\textbf{Output.} The final binary design gets the dense check (S4), giving \(c_{\mathrm{open}}\), \(c_{03}\) and the extremal k-set, which the MPB claim reuses. The check is repeated at cutoff 2.3 \(k_{n}\). Every L-BFGS iterate is stored (coefficients, J, smooth-design gap). Every third iterate and every round end are also stored as binary designs with bands n \ensuremath{-} 4 \ldots{} n + 5 on kgrid(4): these are the 84 092 designs of S1, for surrogate training. A v2 case takes about 6 h on 4 cores; an r6b case up to 16 h.

\textbf{Starts compared}. Among starts that fill the shell evenly, Riesz-weighted shells give the lowest contrast in all three cells, but by only 0.02--0.03 over the next-best start, with 1--3 seeds each: suggestive, not conclusive. Polyhedral starts do better on bcc 136 and sc 88 (main text, Fig. 1). Best dense \(c_{03}\) in plane waves, all with the same protocol (cylinders from \texttt{p3d\_gopt3d\_cyl}, the rest from \texttt{p3d\_gopt3d\_v2}):

\begin{table}[!htbp]
\renewcommand{\thetable}{S3b}
\caption{Best dense plane-wave \(c_{03}\) by kind of start, all with the same protocol.}
\centering\small
\begin{tabularx}{\linewidth}{l>{\raggedright\arraybackslash\hsize=1.12\hsize}X>{\raggedright\arraybackslash\hsize=0.97\hsize}Xc>{\raggedright\arraybackslash\hsize=0.90\hsize}X}
\hline\hline
Lattice, n & Riesz-weighted shell (3 seeds) & Equal-weight shell (1 seed) & Random (2 seeds) & Cylindrical waves, 6 icosahedral axes (1 seed) \\
\hline
bcc 136 & 2.790 & 2.868 & 2.885 & 2.805 \\
fcc 118 & 2.947 & 3.223 & 2.972 & 3.063 \\
sc 88 & 3.140 & 3.163 & 3.224 & 3.506 \\
\hline\hline
\end{tabularx}
\end{table}

Among cylindrical-wave sums, axes along the 6 icosahedral five-fold directions work best and cube-aligned axes worst (S8).

\section*{S4. Finding the band edges}

Free-form designs have no point symmetry, so their band extrema can sit anywhere in the Brillouin zone. An optimizer that sees a fixed set of k-points learns to put them in between. Every check therefore combines two grids, random points and a continuous search for the extrema.

\textbf{Grids.} kgrid(g) is the g \ensuremath{\times} g \ensuremath{\times} g grid \(t_{i}\) = j/g \ensuremath{-} 1/2 in fractional reciprocal coordinates, reduced only by time reversal (k \textasciitilde{} \ensuremath{-}k).

\begin{table}[!htbp]
\renewcommand{\thetable}{S4}
\caption{k-point grids and sets of the band-edge search.}
\centering\small
\begin{tabularx}{\linewidth}{>{\raggedright\arraybackslash\hsize=1.15\hsize}Xc>{\raggedright\arraybackslash\hsize=0.85\hsize}X}
\hline\hline
Grid or set & k-points & Contains \\
\hline
kgrid(4) & 36 & the 2-grid \\
kgrid(6) & 112 & the 2- and 3-grids \\
kgrid(8) & 260 & the 2- and 4-grids \\
kgrid(12) & 868 & the 2-, 3-, 4- and 6-grids, not the 8-grid \\
kgrid(8) \ensuremath{\cup} kgrid(12) + 400 random (plane-wave dense check) & 1492 & \textasciitilde{} \\
kgrid(6) \ensuremath{\cup} kgrid(8) + extremal set (MPB claim scan) & 433--470 & \textasciitilde{} \\
\hline\hline
\end{tabularx}
\end{table}

\textbf{Refinement.} From the 12 highest points of band n and the 12 lowest of band n + 1, Nelder--Mead in k (unconstrained, folded back into the zone; initial simplex 0.03, tolerances 10\(^{-3}\) in k and 10\(^{-7}\) in frequency, at most 60 evaluations) maximizes band n or minimizes band n + 1. The extremal set is the refined points plus the 20 worst raw points per edge.

\textbf{Contrast search.} On the extremal set, c is raised in steps of 4 \% until the gap reaches the target, then bisected: \(c_{\mathrm{open}}\) for a gap \textgreater{} 10\(^{-5}\), \(c_{03}\) for 0.3 \%. The extrema move with c, so the scan and refinement are repeated at \(c_{03}\) and the search is redone on the union of both extremal sets.

\textbf{Why: the withdrawn first claims.} Our first designs were optimized and checked on kgrid(4), kgrid(8) and random points. Two of them seemed to open gaps at index 1.578 (bcc 136, c = 2.49) and 1.594 (fcc 118, c = 2.54). Dense checks closed both:

\begin{itemize}
\tightlist
\item
  \textbf{fcc 118} (design \texttt{p20\_s3}) closes on the line (\ensuremath{-}t, 0, \ensuremath{-}t) at t = 1/3 and t = 1/6. These points are on the 6-grid but not on the 4- or 8-grid. MPB at resolution 48 and c = 2.644 confirms it: gap \ensuremath{-}0.007 \%.
\item
  \textbf{bcc 136} (design \texttt{p0\_s0}) at its design contrast 2.4857: +0.32 \% on the 6- and 8-grids, \ensuremath{-}0.06 \% on the 12-grid, \ensuremath{-}0.23 \% after refinement.
\end{itemize}

Re-checking all 92 earlier designs this way (\texttt{p3d\_recheck3d}) moved the best values to \(c_{03}\) = 3.03 (fcc 118), 3.16 (sc 88), 3.29 (bcc 118) and 3.38 (bcc 136). The redesign with refined edges inside the loop (S3, from v2 on) brought bcc 136 back down to 2.79, a value that survives every later check; polyhedral starts and two push rounds then reached 2.44 (S12).

\section*{S5. Numerical validation and whole-zone coverage of a finite plane-wave model}

A claim states that the binary structure at contrast c has \(\max_{k}\) \(\ensuremath{\omega}_{n}\)(k) \textless{} \(\min_{k}\) \(\ensuremath{\omega}_{n+1}\)(k) over the whole Brillouin zone, for both polarizations. We establish it with MPB through four controlled steps: a k-scan, a continuous search for the band edges, a resolution series and an extrapolation. This is strong numerical evidence, not a computer-assisted proof; the limits are listed at the end.

\textbf{What is claimed.} The gap ratio is

\begin{equation}
\Delta = \frac{2\,(\min_k \omega_{n+1} - \max_k \omega_n)}{\min_k \omega_{n+1} + \max_k \omega_n}
\end{equation}

for \ensuremath{\varepsilon} = c inside the level set \{g/\(g_{\mathrm{rms}}\) \textgreater{} t\} and 1 outside. The set is passed to MPB \cite{JohnsonJoannopoulos2001} as a material function, so MPB samples the continuous surface with its subpixel averaging rather than a fixed voxel image. The primitive cell has unit volume, as in S2.

\textbf{Solver settings.}

\begin{itemize}
\tightlist
\item
  \textbf{Bands:} n + 10, so the gap bands are far from the top of the run, where MPB\textquotesingle s bands converge poorly.
\item
  \textbf{Tolerance:} 10\(^{-8}\).
\item
  \textbf{Block size:} MPB solves bands in blocks and tests convergence on each block\textquotesingle s eigenvalue sum, so a block\textquotesingle s edge bands converge worst. The block size B is the smallest B \ensuremath{\geq} 12 that leaves at least 3 bands of the same block on each side of the gap: 12 for bcc 136 and sc 88, 14 for fcc 118.
\item
  \textbf{Parallelism:} single-threaded MPB processes, each solving its share of the k-points. Memory is about 0.45 GB \ensuremath{\times} (res/32)\(^{3}\) \ensuremath{\times} (bands/142) per process, about 7 GB at res 80.
\end{itemize}

\textbf{The protocol} (\texttt{mpb3d\_proof.py}; one claim takes 4--29 h on 24--32 cores):

\begin{enumerate}
\def\labelenumi{\arabic{enumi}.}
\tightlist
\item
  \textbf{Entry contrast.} MPB at resolution 40 on the plane-wave extremal set (S4) plus the 24 plane-wave worst points, at c = 0.98, 1.02, \ldots, 1.2 \ensuremath{\times} \(c_{03}\), upward until the gap there reaches 0.1 \%. The first passing contrast is the claim contrast.
\item
  \textbf{Scan.} All bands at 433--470 k-points, kgrid(6) \ensuremath{\cup} kgrid(8) \ensuremath{\cup} the extremal set, at resolution 40. The uniform part also gives the density of states.
\item
  \textbf{Continuous edge search.} From the 10 highest points of band n and the 10 lowest of band n + 1, Nelder--Mead in k with MPB itself (\ensuremath{\leq} 40 evaluations each, about 820 in total). Each search warm-starts one solver from its previous k. Every evaluation is recorded, and the gap is taken over all of them and the scan.
\item
  \textbf{Resolution series.} The 4 best refined maxima and minima plus the 2 best grid points of each edge (\ensuremath{\leq} 12 k-points), cold-started at resolutions 48, 64 and 80, and extrapolated linearly in 1/res\(^{2}\) (and in 1/res as a check).
\end{enumerate}

\textbf{Acceptance.} A gap is claimed only if all of these hold:

\begin{enumerate}
\def\labelenumi{\arabic{enumi}.}
\tightlist
\item
  the gap after step 3 is positive;
\item
  the gap is positive at every resolution of step 4;
\item
  the gap does not shrink as the resolution increases;
\item
  the gap is at least three times the sum of the refinement loss (scan gap minus refined gap) and the spread between the two extrapolations (1/res\(^{2}\) and 1/res). For the design at index 1.547 this is 0.252 against 3 \ensuremath{\times} 0.034. The margin is empirical: the two terms are neither independent statistical errors nor guaranteed bounds on unsampled extrema or on the discretization error. Table S6 gives the threshold of every run; designs that fail it are listed there as candidates.
\end{enumerate}

The quoted value is the step-3 gap at resolution 40, the most conservative number; the extrapolated gap is reported next to it.

\begin{table}[!htbp]
\renewcommand{\thetable}{S5}
\caption{Error sources of a claim, how each is controlled, and the evidence.}
\centering\small
\begin{tabularx}{\linewidth}{>{\raggedright\arraybackslash\hsize=0.73\hsize}X>{\raggedright\arraybackslash\hsize=0.96\hsize}X>{\raggedright\arraybackslash\hsize=1.30\hsize}X}
\hline\hline
Error source & Control & Evidence \\
\hline
Band edge between sampled k & two grids, plane-wave and MPB extremal points, continuous search in both solvers & refinement lowers the gap by 0.02--0.07 percentage points (S6) and closes none of the claims \\
Eigensolver convergence & tolerance 10\(^{-8}\), 10 guard bands, gap kept inside a block & at res 32, (n + 6, 10\(^{-5}\)), (n + 10, 10\(^{-7}\)) and (n + 14, 10\(^{-8}\)) agree to 3 \ensuremath{\times} 10\(^{-5}\) in frequency \\
Discretization & resolution 40 \ensuremath{\rightarrow} 80 and extrapolation & the gap grows with resolution in every claim; on two earlier designs it converges smoothly up to res 96 (sc 88: 0.387 \ensuremath{\rightarrow} 0.452 \%; fcc 118: 0.301 \ensuremath{\rightarrow} 0.394 \%) \\
Representation of the structure & continuous level set with subpixel averaging, not a voxel image & plane waves on the voxelized design give smaller gaps (S2) \\
Model & linear, isotropic, lossless two-phase dielectric & standard; dispersion and loss are outside the claim \\
\hline\hline
\end{tabularx}
\end{table}

\textbf{Opening contrast} (\texttt{mpb3d\_open.py}). On the claim\textquotesingle s edge set plus the 20 + 20 most extreme scan points, at resolution 48, c is lowered by 3 \% per step until the gap closes, then bisected twice. This is an estimate on a fixed k-set: at lower c other k-points can become extremal. An opening value becomes a claim only after the full protocol runs at that contrast (for bcc 136 at c = 2.55 this gave a candidate, S6).

\textbf{What is not rigorous.}

\begin{itemize}
\tightlist
\item
  \textbf{Edge-search starts.} Nelder--Mead starts from the 10 highest and 10 lowest scan points. A maximum formed at an avoided crossing is about gap/(2v) \ensuremath{\approx} 0.01 wide in k, while the grids leave gaps of up to about 0.07, so the nearest scan point can sit below the top 10 and the extremum can be missed. Closed for the Riesz-start bcc 136 design (index 1.654; campaign \texttt{p3d\_mpb3d\_verify}): plane-wave gradient searches with Hellmann--Feynman group velocities from 1596 starts (the 1492 scan points, the vertices and face centres of the Bragg polyhedron, the claim\textquotesingle s edges) found 322 and 304 distinct extrema; MPB polished the 24 + 24 most extreme and the claim\textquotesingle s 12 edge points at resolution 40 with Newton steps on its own group velocities. The global gap is 0.4283 \% against the claim\textquotesingle s 0.4284 \%. The v3 candidate (index 1.587): 0.1691 \% against 0.1691 \%. bcc 136 v4 (index 1.584): from 1620 starts, 0.2498 \% against the claim\textquotesingle s 0.2499 \%; re-polished 0.2861, 0.3269 and 0.3429 \% at resolutions 48, 64 and 80 (claim 0.286, 0.327 and 0.343 \%), unchanged at tolerances 10\(^{-9}\) and 10\(^{-10}\), and a second block size moves bands n and n + 1 by 3 \ensuremath{\times} 10\(^{-10}\). Resolutions 96 and 112: 0.3513 and 0.3567 \%; extrapolated 0.371 \% (1/res\(^{2}\)) and 0.391 \% (1/res). The search polished only the 24 + 24 most extreme of 301 / 280 distinct plane-wave extrema, and the plane-wave models rank extrema differently by about 0.2 gap widths: the band n + 1 minimum of the certified model (below) ranks 50th in the design model. A probe therefore evaluated 178 further candidates in MPB at resolution 40 (every unpolished design-model extremum within 0.5 gap widths, the certified model\textquotesingle s extrema and its local-search extrema; \texttt{p3d\_mpb3d\_probe\_v4}) and polished the 87 that lie within half a gap width of an edge: none passes the claim\textquotesingle s edges (gap 0.2498 \%), and near the certified model\textquotesingle s minimum MPB has a band n + 1 minimum 0.36 gap widths above its bottom edge. The polyhedral-start designs were searched the same way, with every distinct extremum within 0.6 gap widths of an edge polished, which includes the probe: index 1.547, 0.2515 \% (claim 0.2515 \%; 0.2787 / 0.3007 / 0.3078 \% at resolutions 48 / 64 / 80); index 1.565, 0.2839 \% (claim 0.2941 \%; 0.2888 / 0.3021 / 0.3069 \%, still 2.2 times the threshold); sc 88 at index 1.681, 0.4242 \% (claim 0.4243 \%).
\item
  \textbf{Fixed edge set across resolutions.} The resolution series reuses the k-points found at resolution 40. Extrema move with resolution, so a fixed set can only overstate the gap at higher resolution. Closed for the design at 1.654: re-polished with Newton steps at every resolution, the gap is 0.4488 / 0.4636 / 0.4747 \% at resolutions 48 / 64 / 80 (0.4488 / 0.4638 / 0.4750 \% on the fixed set), extrapolated 0.488 \% (1/res\(^{2}\)) and 0.513 \% (1/res). Resolution 96: 0.4787 \%. Resolution 112: 0.4815 \%.
\item
  \textbf{Eigensolver tolerance.} MPB tests convergence on a block\textquotesingle s eigenvalue sum, not on each band, and its Ritz values lie above the true ones, so an unconverged band n + 1 inflates the gap. Closed for the design at 1.654: re-solved warm at tolerances 10\(^{-9}\) and 10\(^{-10}\), the gap is unchanged to four digits at resolutions 48, 64 and 80, and a second block size moves bands n and n + 1 by at most 2 \ensuremath{\times} 10\(^{-10}\) (the v3 design: 3 \ensuremath{\times} 10\(^{-10}\)).
\item
  \textbf{Discretization.} The resolution error is extrapolated, not bounded.
\item
  \textbf{Toward a proof.} A computer-assisted proof needs rigorous eigenvalue enclosures over the whole zone, as done for 2D structures \cite{HoangPlumWieners2009}. An intermediate step is a numerical covering certificate for a finite plane-wave model (certify3d.py; floating-point arithmetic, without interval enclosures). On one fixed set of plane waves for the whole zone the inverse-dielectric matrix \ensuremath{\eta} does not depend on k, and every band moves in k at most at speed L = \(\ensuremath{\lambda}_{\mathrm{max}}\)(\ensuremath{\eta})\(^{1/2}\) (0.997 for the Ho model of bcc 136). Cartesian cubes cover half the Wigner--Seitz cell; a cube passes when, at its centre, either this speed limit or a second-order bound from the eigenvectors of the bands next to the gap keeps band n below and band n + 1 above a chosen interval inside the gap. Pilots on bcc 136 pass cubes of side 0.01 at the band edges, where the speed limit alone needs 0.0025. Certificates are complete for eight designs, each with no unresolved cube and no violation. At 0.9 of the model gap: bcc 136 at index 1.654 (156 419 cubes, 8.2 h on 32 cores; certified model gap 0.335 \%), fcc 118 at 1.680 (164 887 cubes, 5.0 h; 0.272 \%), sc 88 at 1.759 (87 610 cubes, 1.7 h; 0.223 \%), fcc 118 at 1.677 (159 481 cubes, 4.3 h; 0.306 \%), bcc 136 at 1.565 (206 329 cubes, 12.4 h; 0.292 \%), bcc 136 at 1.547 (204 623 cubes, 11.7 h; 0.277 \%) and sc 88 at 1.681 (167 118 cubes, 3.6 h; 0.355 \%). For bcc 136 at 1.584 the first run stopped at 84 cubes, all inside the cell, where band n + 1 of the model dips to 2.03462 near k = (0.279, \ensuremath{-}0.206, \ensuremath{-}0.050) (MPB coordinates), below the 2.03514 found from the claim\textquotesingle s edges: the model gap is 0.345 \%, not 0.371 \%. Resumed with the corrected edges at 0.8 of the model gap, keeping the cubes already certified (valid for any narrower interval), it completes (285 865 cubes, 13.8 h in three runs; certified model gap 0.306 \%). MPB does not follow the model there (edge-search starts, above). The certificate is a statement about the model. For bcc 136 the certified model has a 0.373 \% gap over the 442 scan points and 0.412 \% at the 12 edge points (MPB 0.443 \% at resolution 40, 0.475 \% at 80); after a uniform 1.2 \% shift its bands n and n + 1 differ from MPB\textquotesingle s by 0.12 MPB gap widths rms. Within one model gap of the edges MPB lies at most 0.19 (band n) and 0.30 (band n + 1) gap widths further into the gap than the model; an MPB state at mid-gap would need 0.38 against a certificate of 0.9 of the model gap. For the two-round bcc 136 design (index 1.584) the model gap is 0.345 \% (0.371 \% over the MPB scan, which misses the dip; MPB 0.264 \% at resolution 40, 0.343 \% at 80), the residual is 0.19 widths rms, and the excursions are 0.41 and 0.46 widths over the scan against 0.52 needed for a certificate of 0.8 of the model gap (0.33 against 0.40 at the edges, resolution 80); at the dip MPB\textquotesingle s band n + 1 lies 0.36 gap widths above its own bottom edge, so there the model errs toward a smaller gap; for fcc 118 (index 1.677) 0.340 \% (0.344 \% over the scan; MPB 0.432 \%), 0.15 rms, 0.22 and 0.351 against 0.355 (0.14 against 0.32 at resolution 80) (model\_check3d.py) (S2). The certificates hold for the plane-wave models in floating-point arithmetic; they are not proofs for Maxwell\textquotesingle s equations, and the only rigorous statement remains the lower bound of S7.
\end{itemize}

\textbf{Second-order cube test} (certify3d.py). At a cube centre, Y and Z hold the eigenvectors of bands n \ensuremath{-} p + 1 \ldots{} n and n + 1 \ldots{} n + p of the fixed-set operator M(k), with eigenvalues \(\ensuremath{\Lambda}_{Y}\) and \(\ensuremath{\Lambda}_{Z}\). The second-order part of M in the displacement \ensuremath{\delta} from the centre is positive semidefinite and at most L\(^{2}\)\textbar \ensuremath{\delta}\textbar{}\(^{2}\); the smallest eigenvalue of the affine block \(\ensuremath{\Lambda}_{Z}\) + \ensuremath{\delta}\ensuremath{\cdot}B is concave and the largest of \(\ensuremath{\Lambda}_{Y}\) + \ensuremath{\delta}\ensuremath{\cdot}B\ensuremath{'} convex in \ensuremath{\delta}, so their extremes over the cube lie at its 8 corners. The bands outside the two clusters are held off by the speed limit, and a Schur complement with Courant--Fischer gives, everywhere in a cube of circumradius r,

\begin{equation}
\begin{aligned}
&M(k) = C(k)^{\dagger}\,\eta\,C(k), \qquad B_a = Z^{\dagger}\,\partial_{k_a}M\,Z, \qquad B'_a = Y^{\dagger}\,\partial_{k_a}M\,Y,\\
&a_Z^2 = \textstyle\sum_a \|\partial_{k_a}M\,Z\|^2, \qquad a_Y^2 = \textstyle\sum_a \|\partial_{k_a}M\,Y\|^2,\\
&\omega_{n+1}^2 \ge \tfrac12\Bigl[m+f-\sqrt{(f-m)^2+4r^2a_Z^2}\,\Bigr], \qquad m = \min_{\rm corners}\lambda_{\min}(\Lambda_Z+\delta\cdot B), \qquad f = \omega_{n+p+1}^2 - 2Lr\,\omega_{n+p+1},\\
&\omega_n^2 \le \tfrac12\Bigl[m'+f'+\sqrt{(m'-f')^2+4c'^2}\,\Bigr], \qquad m' = \max_{\rm corners}\lambda_{\max}(\Lambda_Y+\delta\cdot B') + L^2r^2,\\
&\qquad f' = (\omega_{n-p}+Lr)^2, \qquad c' = r\,a_Y + L^2 r^2 ,
\end{aligned}
\end{equation}

with the frequencies \ensuremath{\omega} on the right taken at the centre. A cube passes for band n when this bound or the speed limit \(\ensuremath{\omega}_{n}\)(\(k_{c}\)) + Lr lies below the target interval, and for band n + 1 likewise.

\section*{S6. Verification data per design}

Eighteen designs pass every acceptance test of S5, including the claim threshold; eight more are candidates, with a positive gap at every resolution but short of the threshold (Table S6a). The table has 19 accepted rows for 18 geometries (the one-push icosidodecahedral design at two contrasts) and ten candidate rows, two of them lower-contrast probes of accepted geometries. From the icosidodecahedral start (S12) the design passes at index 1.572 as it stands, at 1.565 and 1.555 after one round of perturbed restarts, and at 1.547 after a second. Pushing on from the Riesz-start designs (\texttt{p3d\_gopt3d\_v3} and v4, perturbed restarts from the best designs) lowers the contrast at which MPB confirms a gap: bcc 136 from index 1.654 to 1.587 (v3, a candidate: positive at every resolution but short of the threshold, mostly through the spread of the two extrapolations) and 1.584 (v4, passes); fcc 118 from 1.699 (a candidate) to 1.680 (v3) and 1.677 (v4). The Riesz-start bcc 136 design at the fixed contrast c = 2.55 (index 1.597) has a positive gap at every resolution (0.103 \% refined, 0.154 \% at resolution 80) but fails the threshold. In every design the refined gap grows by 0.03--0.09 percentage points from resolution 40 to 80.

\begin{table}[!htbp]
\renewcommand{\thetable}{S6a}
\caption{MPB verification of every claimed design and every candidate.}
\centering\small
\begin{tabularx}{\linewidth}{>{\raggedright\arraybackslash\hsize=1.17\hsize}Xccc>{\raggedright\arraybackslash\hsize=1.12\hsize}X>{\raggedright\arraybackslash\hsize=0.71\hsize}Xc}
\hline\hline
Design & c (index) & Scan & Refined & Res 48 / 64 / 80 & Extrapolated, 1/res\(^{2}\) (1/res) & Threshold \\
\hline
bcc 136 ID, two pushes & 2.3925 (1.5468) & 0.257 & 0.252 & 0.279 / 0.301 / 0.308 & 0.326 (0.355) & 0.103 \\
bcc 136 ID, one push, lower c & 2.4180 (1.5550) & 0.275 & 0.260 & 0.277 / 0.292 / 0.301 & 0.313 (0.336) & 0.113 \\
bcc 136 ID, one push & 2.4487 (1.5648) & 0.310 & 0.284 & 0.289 / 0.302 / 0.307 & 0.318 (0.335) & 0.132 \\
bcc 136 ID & 2.4705 (1.5718) & 0.234 & 0.231 & 0.252 / 0.271 / 0.281 & 0.297 (0.325) & 0.091 \\
bcc 136 ID with background & 2.4964 (1.5800) & 0.151 & 0.136 & 0.154 / 0.175 / 0.183 & 0.201 (0.230) & 0.133 \\
bcc 136 Riesz, two pushes & 2.5084 (1.5838) & 0.264 & 0.250 & 0.286 / 0.327 / 0.343 & 0.377 (0.432) & 0.210 \\
bcc 136 IDIC (symmetric), one push & 2.5402 (1.5938) & 0.237 & 0.221 & 0.244 / 0.271 / 0.283 & 0.305 (0.342) & 0.159 \\
bcc 136 IDIC (random), one push & 2.6050 (1.6140) & 0.234 & 0.226 & 0.244 / 0.266 / 0.278 & 0.297 (0.330) & 0.125 \\
bcc 136 IDIC (symmetric) & 2.6496 (1.6278) & 0.279 & 0.249 & 0.268 / 0.291 / 0.298 & 0.316 (0.345) & 0.178 \\
bcc 136 T42, one push & 2.7203 (1.6493) & 0.391 & 0.346 & 0.378 / 0.413 / 0.428 & 0.456 (0.504) & 0.280 \\
bcc 136 Riesz & 2.7343 (1.6536) & 0.443 & 0.428 & 0.449 / 0.464 / 0.475 & 0.488 (0.513) & 0.120 \\
bcc 136 S34, one push & 2.7604 (1.6614) & 0.454 & 0.437 & 0.454 / 0.467 / 0.474 & 0.486 (0.506) & 0.111 \\
bcc 136 S34 & 2.8052 (1.6749) & 0.440 & 0.432 & 0.461 / 0.482 / 0.491 & 0.508 (0.537) & 0.113 \\
fcc 118 Riesz, two pushes & 2.8108 (1.6765) & 0.432 & 0.409 & 0.436 / 0.465 / 0.479 & 0.503 (0.545) & 0.196 \\
fcc 118 Riesz, one push & 2.8222 (1.6799) & 0.383 & 0.360 & 0.389 / 0.410 / 0.423 & 0.440 (0.473) & 0.168 \\
sc 88 A, one push & 2.8272 (1.6814) & 0.444 & 0.424 & 0.448 / 0.478 / 0.491 & 0.515 (0.556) & 0.183 \\
sc 88 A & 2.8471 (1.6873) & 0.558 & 0.493 & 0.511 / 0.536 / 0.545 & 0.565 (0.598) & 0.295 \\
fcc 118 random & 2.9735 (1.7244) & 0.403 & 0.332 & 0.348 / 0.364 / 0.368 & 0.381 (0.401) & 0.272 \\
sc 88 random & 3.0926 (1.7586) & 0.349 & 0.304 & 0.313 / 0.332 / 0.343 & 0.358 (0.387) & 0.222 \\
Candidate: bcc 136 ID with background, lower c & 2.4806 (1.5750) & 0.143 & 0.127 & 0.144 / 0.165 / 0.174 & 0.191 (0.220) & 0.137 \\
Candidate: bcc 136 Riesz, one push & 2.5178 (1.5868) & 0.204 & 0.169 & 0.190 / 0.213 / 0.225 & 0.244 (0.277) & 0.207 \\
Candidate: bcc 136 Riesz, lower c & 2.5500 (1.5969) & 0.139 & 0.103 & 0.117 / 0.144 / 0.154 & 0.177 (0.214) & 0.219 \\
Candidate: bcc 208 Riesz, one push & 2.7499 (1.6583) & 0.160 & 0.132 & 0.137 / 0.153 / 0.160 & 0.173 (0.195) & 0.151 \\
Candidate: fcc 118 S51, one push & 2.7757 (1.6660) & 0.294 & 0.225 & 0.259 / 0.297 / 0.314 & 0.345 (0.399) & 0.369 \\
Candidate: bcc 136 IDIC (random) & 2.8071 (1.6754) & 0.234 & 0.183 & 0.210 / 0.236 / 0.248 & 0.269 (0.307) & 0.263 \\
Candidate: fcc 118 Riesz & 2.8877 (1.6993) & 0.177 & 0.108 & 0.153 / 0.183 / 0.196 & 0.221 (0.263) & 0.332 \\
Candidate: fcc 118 S51 & 2.9518 (1.7181) & 0.452 & 0.337 & 0.375 / 0.411 / 0.427 & 0.457 (0.507) & 0.496 \\
Candidate: bcc 118 random & 3.3522 (1.8309) & 0.131 & 0.107 & 0.133 / 0.162 / 0.175 & 0.199 (0.240) & 0.195 \\
Candidate: bcc 190 TD & 3.1049 (1.7621) & 0.096 & 0.050 & 0.078 / 0.098 / 0.108 & 0.125 (0.154) & 0.224 \\
\hline\hline
\end{tabularx}
\end{table}

Gaps in percent: on the scan (433--470 k-points, resolution 40), after the edge search (about 820 MPB evaluations), on the edge set at resolutions 48, 64 and 80, extrapolated, and the claim threshold of S5. The index-1.565 row uses the band edges of the global search (S5); its claim run gave 0.294 \% refined and 0.310 / 0.325 / 0.333 \% at resolutions 48 / 64 / 80. ID, IDIC, S34, S51 and A are polyhedral starts (S12; IDIC with random or symmetric signs), Riesz the log-Riesz-weighted shell (S3); a push is one round of perturbed restarts; lower c means the same design at a lower contrast. Each claim took 4--29 h on 24--32 cores.

Entry contrasts: each design passed the 0.1 \% entry test at the first rung, 0.98 \(c_{03}\), except bcc 118, which needed the second rung (0.023 \% at 3.2208, then 0.131 \% at 3.3522). The edge k-points, the bands n \ensuremath{-} 2 \ldots{} n + 3 at every resolution, the full scan spectra and every Nelder--Mead evaluation are in the claim files (S10).

\begin{table}[!htbp]
\renewcommand{\thetable}{S6c}
\caption{Band edges of the design at index 1.547 (MPB, c = 2.3925, 146 bands, tolerance 10\(^{-8}\)): the 12 edge k-points of the claim, all from the continuous edge search, and the frequency (MPB units, cell volume 1) of the band edge each one carries, at resolution 40 in the edge search and on the cold-started resolution series. The global search of S5 (1620 starts, every distinct extremum within 0.6 gap widths of an edge re-polished in MPB) finds no point beyond these edges: 0.2515 \% against the claim\textquotesingle s 0.2515 \% at resolution 40.}
\centering\small
\begin{tabular}{lccccc}
\hline\hline
k (reciprocal-lattice coordinates) & Edge & Res 40 & Res 48 & Res 64 & Res 80 \\
\hline
(\ensuremath{-}0.3477, \ensuremath{-}0.2210, +0.2567) & max of band n & 2.07488 & 2.07398 & 2.07302 & 2.07261 \\
(\ensuremath{-}0.3465, \ensuremath{-}0.2319, +0.2666) & max of band n & 2.07479 & 2.07388 & 2.07290 & 2.07248 \\
(\ensuremath{-}0.3461, \ensuremath{-}0.2225, +0.2578) & max of band n & 2.07488 & 2.07397 & 2.07302 & 2.07261 \\
(\ensuremath{-}0.3461, \ensuremath{-}0.2225, +0.2571) & max of band n & 2.07488 & 2.07397 & 2.07302 & 2.07261 \\
(\ensuremath{-}0.3457, \ensuremath{-}0.2350, +0.2702) & max of band n & 2.07471 & 2.07380 & 2.07282 & 2.07240 \\
(\ensuremath{-}0.2752, +0.2267, \ensuremath{-}0.2640) & max of band n & 2.07460 & 2.07354 & 2.07259 & 2.07215 \\
(\ensuremath{-}0.3963, \ensuremath{-}0.3957, +0.3237) & min of band n + 1 & 2.08039 & 2.07979 & 2.07927 & 2.07901 \\
(\ensuremath{-}0.0554, \ensuremath{-}0.4798, \ensuremath{-}0.4593) & min of band n + 1 & 2.08020 & 2.07983 & 2.07946 & 2.07930 \\
(\ensuremath{-}0.0471, \ensuremath{-}0.4754, \ensuremath{-}0.4565) & min of band n + 1 & 2.08010 & 2.07977 & 2.07942 & 2.07928 \\
(\ensuremath{-}0.0471, \ensuremath{-}0.4746, \ensuremath{-}0.4564) & min of band n + 1 & 2.08010 & 2.07977 & 2.07943 & 2.07929 \\
(\ensuremath{-}0.0457, \ensuremath{-}0.4754, \ensuremath{-}0.4580) & min of band n + 1 & 2.08011 & 2.07977 & 2.07942 & 2.07928 \\
(\ensuremath{-}0.0398, \ensuremath{-}0.4780, \ensuremath{-}0.4567) & min of band n + 1 & 2.08013 & 2.07980 & 2.07944 & 2.07930 \\
Gap (\%) & \textasciitilde{} & 0.252 & 0.279 & 0.301 & 0.308 \\
\hline\hline
\end{tabular}
\end{table}

\begin{table}[!htbp]
\renewcommand{\thetable}{S6b}
\caption{Opening searches for five earlier designs: resolution 48, edge set; c lowered 3 \% per step from 0.97 c, then bisected.}
\centering\small
\begin{tabular}{lccc}
\hline\hline
Design & Gap still open at c (index) & Closed at c & Gap at the last open step \\
\hline
bcc 136, Riesz s0 & 2.4766 (1.574) & 2.4578 & 0.014 \% \\
fcc 118, random s1 & 2.6933 (1.641) & 2.6728 & 0.003 \% \\
fcc 118, Riesz s1 & 2.8225 (1.680) & 2.8011 & 0.030 \% \\
sc 88, random s2 & 2.8441 (1.686) & 2.8225 & 0.019 \% \\
bcc 118, warm s6 & 3.2025 (1.790) & 3.1782 & 0.016 \% \\
\hline\hline
\end{tabular}
\end{table}

The gap falls almost linearly with contrast in every design, at about 1.8 percentage points per unit of c for bcc 136, so the opening contrasts are well defined on these k-sets.

\begin{figure}[tb]
\includegraphics[width=\linewidth]{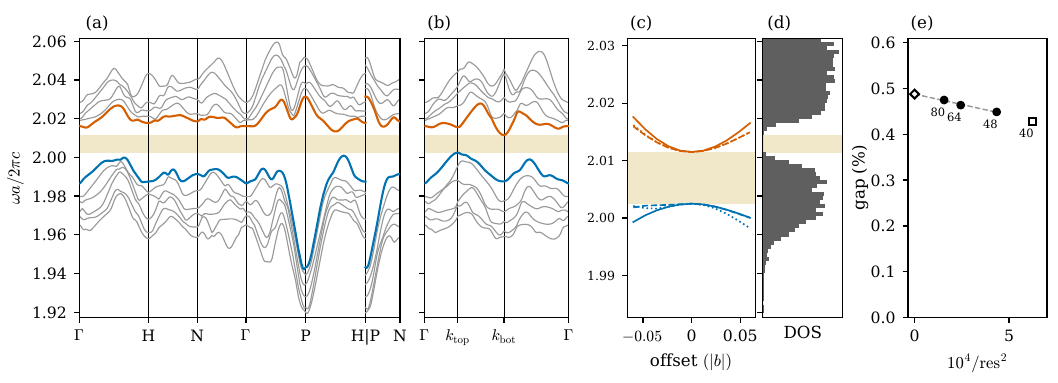}
\caption{Band structure of the Riesz-start bcc 136 design at c = 2.7343 (index 1.654). (a) Bands n \ensuremath{-} 5 \ldots{} n + 5 along \ensuremath{\Gamma}--H--N--\ensuremath{\Gamma}--P--H\textbar P--N; the extrema of bands n and n + 1 lie off the high-symmetry lines, so the gap on this path (0.63 \%) overstates the complete gap. (b) The path \ensuremath{\Gamma} \ensuremath{\rightarrow} \(k_{\mathrm{top}}\) \ensuremath{\rightarrow} \(k_{\mathrm{bottom}}\) \ensuremath{\rightarrow} \ensuremath{\Gamma} through the refined band edges, and (c) six cuts of \ensuremath{\pm}0.06 \textbar b\textbar{} through them, on an expanded scale; the gap on both is 0.449 \% at resolution 48, the claim\textquotesingle s value. (d) Density of states on the 12\(^{3}\) grid (gap 0.48 \% at resolution 40). (e) Gap against 1/res\(^{2}\). The band structure of the design at index 1.547 is Fig. 2 of the Letter.}
\label{fig:S1}
\end{figure}

\section*{S7. The empty-lattice bound in 3D}

No dielectric structure with \ensuremath{\varepsilon} between 1 and c can have a gap above band n unless c exceeds a bound computed from the empty lattice alone. The bound is rigorous, and it tells us which bands to design for. The verified designs sit 12--16 times above it in c \ensuremath{-} 1.

\textbf{Statement.} Let \(\ensuremath{\nu}_{j}\)(k) be the j-th smallest of the values \textbar k + G\textbar{} over all reciprocal vectors G, each counted twice (two polarizations). If \ensuremath{\varepsilon}(r) takes values in {[}1, c{]} (two-phase or graded, any geometry), a complete gap of ratio \ensuremath{\Delta} above band n requires

\begin{equation}
\sqrt{c} \;\ge\; R_n\,\frac{1+\Delta/2}{1-\Delta/2}, \qquad R_n = \frac{\max_k \nu_n(k)}{\min_k \nu_{n+1}(k)}
\end{equation}

\textbf{Proof.} In the H form, \(\ensuremath{\omega}_{j}\)(k)\(^{2}\) is the j-th min--max value of the Rayleigh quotient \ensuremath{\int} \ensuremath{\varepsilon}\(^{-1}\) \textbar\ensuremath{\nabla} \ensuremath{\times} H\textbar{}\(^{2}\) / \ensuremath{\int} \textbar H\textbar{}\(^{2}\) over divergence-free Bloch fields. The quotient decreases when \ensuremath{\varepsilon} increases pointwise, so every band does too. Comparing with the uniform media \ensuremath{\varepsilon} = c and \ensuremath{\varepsilon} = 1 gives \(\ensuremath{\nu}_{j}\)(k)/\ensuremath{\surd}c \ensuremath{\leq} \(\ensuremath{\omega}_{j}\)(k) \ensuremath{\leq} \(\ensuremath{\nu}_{j}\)(k) for every j and k. Then max \(\ensuremath{\omega}_{n}\) \ensuremath{\geq} max \(\ensuremath{\nu}_{n}\)/\ensuremath{\surd}c and min \(\ensuremath{\omega}_{n+1}\) \ensuremath{\leq} min \(\ensuremath{\nu}_{n+1}\). The gap needs min \(\ensuremath{\omega}_{n+1}\) / max \(\ensuremath{\omega}_{n}\) = (1 + \ensuremath{\Delta}/2)/(1 \ensuremath{-} \ensuremath{\Delta}/2), which gives the inequality. \ensuremath{\blacksquare}

\textbf{Computation} (\texttt{bounds3d.py}). Unshifted 36\(^{3}\) and 40\(^{3}\) grids in fractional coordinates (they contain \ensuremath{\Gamma}, the zone boundary and the 2-, 3-, 4-, 5-, 6-, 8-, 9-, 10-, 12-, 18- and 20-point grids), 20 000 random points, and Nelder--Mead from the 24 best points of each band, for even n \ensuremath{\leq} 300 and \ensuremath{\Delta} = 0.3 \%. A sampled maximum is never above the true one and a sampled minimum never below, so sampling can only weaken the bound; it stays valid. \(\ensuremath{\nu}_{n}\) has its maxima at cusps, often at \ensuremath{\Gamma} or a high-symmetry point, where Nelder--Mead cannot climb: an earlier version on a grid shifted by half a step missed them and gave values up to 0.02 too low (sc 88: 1.117 instead of 1.132). For odd n the two polarizations of one wave fall on either side of the gap, so the bound is poor, and only even n are useful.

\begin{table}[!htbp]
\renewcommand{\thetable}{S7a}
\caption{Lowest empty-lattice bound \(c_{B}\) in each range of bands (the band in parentheses).}
\centering\small
\begin{tabular}{lcccc}
\hline\hline
Lattice & Lowest bound for n \ensuremath{\leq} 40 & n \ensuremath{\leq} 100 & n \ensuremath{\leq} 200 & n \ensuremath{\leq} 300 \\
\hline
sc & 1.212 (n = 40) & 1.132 (88) & 1.127 (162) & 1.088 (296) \\
fcc & 1.293 (30) & 1.184 (68) & 1.103 (200) & 1.081 (290) \\
bcc & 1.319 (38) & 1.167 (100) & 1.112 (136) & 1.083 (208) \\
hexagonal, ideal c/a & 1.340 (24) & 1.199 (72) & 1.130 (132) & 1.101 (246) \\
\hline\hline
\end{tabular}
\end{table}

The bound is not monotonic in n: sc 56 has 1.46, between sc 40 (1.21) and sc 88 (1.13). We designed only at local minima.

\begin{table}[!htbp]
\renewcommand{\thetable}{S7b}
\caption{The claimed cells against the bound.}
\centering\small
\begin{tabular}{lccc}
\hline\hline
Cell & Bound & Claimed c & (c \ensuremath{-} 1)/(bound \ensuremath{-} 1) \\
\hline
bcc 136, polyhedral start, two push rounds & 1.112 & 2.3925 & 12.4 \\
bcc 136, Riesz start, two push rounds & 1.112 & 2.5084 & 13.5 \\
bcc 136, Riesz start & 1.112 & 2.7343 & 15.4 \\
fcc 118, two push rounds & 1.144 & 2.8108 & 12.6 \\
fcc 118, one push round & 1.144 & 2.8222 & 12.7 \\
fcc 118, random start & 1.144 & 2.9735 & 13.7 \\
sc 88, polyhedral start, one push round & 1.132 & 2.8272 & 13.8 \\
sc 88, random start & 1.132 & 3.0926 & 15.9 \\
\hline\hline
\end{tabular}
\end{table}

In 2D the best designs sit 3--5 times above. The larger factor in 3D reflects what the bound ignores: a two-phase structure cannot push every state of band n down and every state of band n + 1 up at once, and in 3D the vector coupling through a Fourier peak depends on polarization (S8). A tighter, still rigorous bound in the spirit of Rechtsman and Torquato \cite{RechtsmanTorquato2009} is open.

\section*{S8. Reciprocal-space analysis}

The gap-carrying Fourier weight sits on a thin shell at the Bragg sphere. On bcc 136 and fcc 118 the shell weights behave like the charges of a long-range Riesz equilibrium; pruning the weak components destroys the gap; and of the cylindrical-wave starts, the most evenly covering axis set works best.

\textbf{Shell.} Amplitudes \textbar{}\(\hat{\theta}\)(G)\textbar{} of the binary design are measured in units of sin(\ensuremath{\pi}f)/\ensuremath{\pi}, the largest a single component can have at filling f. Of the 218--313 coefficient pairs, 14--21 exceed 0.1 in the Riesz-start designs, on a shell at 0.93--1.0 \ensuremath{\times} 2\(k_{n}\) (about 18 pairs on average).

\textbf{Riesz test} (\texttt{interp/peaks\_riesz.py}).

\begin{enumerate}
\def\labelenumi{\arabic{enumi}.}
\tightlist
\item
  \textbf{Points.} Shell peaks with 0.85 \textless{} \textbar G\textbar/2\(k_{n}\) \textless{} 1.12 and \textbar{}\(\hat{\theta}\)\textbar{} above a threshold \(a_{\mathrm{min}}\) are projected to x = G/\textbar G\textbar{} on the unit sphere. Each comes with its antipode and the same charge.
\item
  \textbf{Force.} For a Riesz s-energy \ensuremath{\Sigma} \(q_{i}\) \(q_{j}\) \textbar{}\(x_{i}\) \ensuremath{-} \(x_{j}\)\textbar{}\(^{\ensuremath{-}s}\) (logarithmic for s = 0), the tangential force on point i is \(q_{i}\) \(f_{i}\), with f = Mq linear in the charges. R(q) = \textbar Mq\textbar{} divided by the force scale; R = 0 is exact equilibrium.
\item
  \textbf{Floor.} Lattice points cannot sit at exact equilibrium positions, so the best achievable value is \(R_{\mathrm{fit}}\), from non-negative least squares on q.
\item
  \textbf{Null model.} The observed weights are permuted over the same positions, 2000 times.
\item
  \textbf{Charge distance.} The relative change that brings the weights halfway to \(R_{\mathrm{fit}}\), against the change permuted weights need (300 draws).
\end{enumerate}

\begin{table}[!htbp]
\renewcommand{\thetable}{S8a}
\caption{Riesz test of the shell peaks, for charges q = \textbar{}\(\hat{\theta}\)\textbar{}\(^{2}\) and logarithmic interaction.}
\centering\small
\begin{tabularx}{\linewidth}{>{\raggedright\arraybackslash\hsize=0.89\hsize}Xc>{\raggedright\arraybackslash\hsize=0.42\hsize}Xcc>{\raggedright\arraybackslash\hsize=1.69\hsize}X}
\hline\hline
Design (\(a_{\mathrm{min}}\), peaks) & R(observed) & R, permuted median & p & \(R_{\mathrm{fit}}\) floor & Change to balance: observed vs permuted \\
\hline
bcc 136 (0.04, 79) & 0.103 & 0.170 & \textless{} 0.0005 & 0.034 & 25 \% vs 44 \% (p \textless{} 0.003) \\
fcc 118 (0.06, 48) & 0.128 & 0.197 & \textless{} 0.0005 & 0.051 & 24 \% vs 47 \% (p \textless{} 0.003) \\
sc 88 (0.04, 72) & 0.153 & 0.182 & 0.12 & 0.021 & 46 \% vs 43 \% (n.s.) \\
\hline\hline
\end{tabularx}
\end{table}

The effect is strongest for the logarithmic energy and fades with shorter range: for bcc 136, p = 0.0055 at s = 1 and 0.12 at s = 2; for fcc 118 it persists to s = 3 (p = 0.012) and fades by s = 6 (p = 0.055).

\textbf{Caveat.} These are the best designs of \texttt{p3d\_gopt3d\_push}, optimized before the dense re-check of S4; under the dense check their gaps need c \ensuremath{\approx} 3--3.4. The finding motivated the Riesz-weighted starts that produced the verified designs.

\textbf{Pruning.} From the Riesz-start designs (P = 313, 268, 218 pairs), keeping the 60, 40 or 25 largest pairs and re-optimizing gives no gap at any contrast tried, in all 9 cases. In 2D, 79 \% of the strong inner peaks are differences of two strong ring vectors, against 57 \% for random vectors (p = 8 \ensuremath{\times} 10\(^{-5}\)).

\textbf{Cylindrical-wave axes.} Dense plane-wave \(c_{03}\) (worse of cutoffs 2.0 and 2.3 \(k_{n}\)) for starts from 3, 4, 6 or 10 axes, against the Riesz-weighted shell:

\begin{table}[!htbp]
\renewcommand{\thetable}{S8b}
\caption{Dense plane-wave \(c_{03}\) of cylindrical-wave starts, by axis set, against the Riesz-weighted shell.}
\centering\small
\begin{tabular}{lccccc}
\hline\hline
Cell & 3 cube axes & 4 cube diagonals & 6 five-fold axes & 10 three-fold axes & Riesz shell \\
\hline
bcc 136 & 3.200 & 2.982 & \textbf{2.805} & 2.999 & 2.790 \\
fcc 118 & no gap & 3.273 & \textbf{3.063} & 3.370 & 2.947 \\
sc 88 & 3.886 & no gap & \textbf{3.506} & no gap & 3.140 \\
\hline\hline
\end{tabular}
\end{table}

The 6 icosahedral five-fold axes, the best packing of 6 lines through a point, win in every cell, and on bcc 136 they come within 0.015 of the Riesz start. One seed per entry. At sc 216, neither cylindrical-wave nor Riesz starts opened a gap at the reduced cutoff 1.7 \(k_{n}\) without contrast escalation.

\section*{S9. The counting estimate}

Coverage, Parseval and the lattice give a first-order count, N \ensuremath{\approx} 2.8 R\(^{3/2}\) shell pairs, and with it a scaling of the Bragg strength b \ensuremath{\propto} n\(^{\ensuremath{-}1/4}\). Like Eq. 5 of \cite{Maiwald2022} it is not a design rule: the designs do not follow it (end of this section). With the amplitudes of a thresholded design, the two-wave version gives c \ensuremath{-} 1 \ensuremath{\gtrsim} 1.7 at fcc 118, close to what the optimizer finds; at the band edges 20--30 plane waves couple, so the two-wave estimate is only a scale (end of this section).

\textbf{1. Coverage.} N pairs give 2N points with their antipodes. Caps of angular radius \(\ensuremath{\theta}_{c}\) around them cover the sphere only if 2N \ensuremath{\cdot} 2\ensuremath{\pi}(1 \ensuremath{-} cos \(\ensuremath{\theta}_{c}\)) \ensuremath{\geq} 4\ensuremath{\pi}, so \(\ensuremath{\theta}_{c}^{2}\)/2 \ensuremath{\geq} 1/N. The best coverings have \(\ensuremath{\theta}_{c}\) \ensuremath{\approx} 1.56/\ensuremath{\surd}N (the six five-fold axes give 1.27/N, the 6 + 15 icosahedral set 1.39/N). A free photon at angle \ensuremath{\theta} from the nearest peak meets that Bragg plane at \textbar k\textbar{} = (R/2)/cos \ensuremath{\theta}, higher by about \ensuremath{\theta}\(^{2}\)/2:

\begin{equation}
\delta \approx \tfrac12\,\theta_c^{2} \approx \frac{1.2}{N}
\end{equation}

\textbf{2. Amplitudes of a thresholded design.} Let \ensuremath{\theta} = 1 where g/\ensuremath{\sigma} \textgreater{} t, with g = \ensuremath{\Sigma} \(\ensuremath{\alpha}_{G}\) exp(2\ensuremath{\pi}i G\ensuremath{\cdot}r), \ensuremath{\sigma}\(^{2}\) = \ensuremath{\Sigma} \textbar{}\(\ensuremath{\alpha}_{G}\)\textbar{}\(^{2}\) and f = 1 \ensuremath{-} \ensuremath{\Phi}(t). If g is close to Gaussian, which a dozen comparable peaks ensure, the first two Hermite terms of the step give

\begin{equation}
\hat\theta(G) \approx \varphi(t)\,\frac{\alpha_G}{\sigma}, \qquad \hat\theta(G_a - G_b) \approx t\,\varphi(t)\,\frac{\alpha_a\,\alpha_b^{*}}{\sigma^{2}}
\end{equation}

with \ensuremath{\varphi} and \ensuremath{\Phi} the standard normal density and distribution. Two consequences: the shell carries the share \ensuremath{\eta} = \ensuremath{\varphi}(t)\(^{2}\)/{[}f(1 \ensuremath{-} f){]} of the Fourier power of the indicator \ensuremath{\theta} (not of the generator), 0.60 at f = 0.35; and thresholding creates inner peaks at differences of shell vectors, weaker by t/\ensuremath{\surd}(2N) per pair for N equal pairs, with a fixed triplet phase in this quadratic (Hermite) term: in phase for f \textless{} 1/2, absent at f = 1/2, reversed for f \textgreater{} 1/2. These are the phase-locked inner peaks of the 2D records.

\textbf{3. Two-wave criterion.} With Bragg strength b = (c \ensuremath{-} 1)/{[}1 + (c \ensuremath{-} 1)f{]}, a single Bragg plane splits free photons by about b\textbar{}\(\hat{\theta}\)\textbar, and a gap would need

\begin{equation}
b\,\varphi(t)\,(2N)^{-1/2} \gtrsim \frac{1.2}{N}
\end{equation}

For N = 18 and f = 0.35 this gives b \ensuremath{\gtrsim} 1.08, or c \ensuremath{-} 1 \ensuremath{\gtrsim} 1.7, close to the 1.6--1.9 the optimizer finds. (With the earlier constants, \ensuremath{\delta} = 0.6/N and a fitted \ensuremath{\gamma} = 0.5 in place of 1/\(\bar{\varepsilon}\), the same criterion gave 2.4.) The band edges are not few-wave vertex states. In the fixed-set Ho model at the 12 refined edge points of bcc 136 at indices 1.584 and 1.654 and fcc 118 at 1.677 (dpt3d.py), the edge state has a participation number of 20--31 plane waves, and 90 \% of its weight lies on 46--60 waves whose free wavenumbers spread over \ensuremath{\pm}8 \% of their mean; only 2--6 free waves lie within 1 \% of it, where a vertex of three planes would put four. Restricted to the nearly degenerate waves, the problem misses the edge by 5--23 gap widths with the waves within 3 \% (10--20 waves) and by 0.3--0.7 with those within 8 \% (35--43 waves); a second-order correction from the remaining waves brings the 3 \% subspace to 0.1--2 gap widths. A closed-form vertex criterion therefore does not apply at these contrasts. Linear response: d\ensuremath{\Delta}/d\(\hat{\theta}\)(G) follows from the edge eigenvectors (Ho rule, d\ensuremath{\eta} = \ensuremath{-}\ensuremath{\eta} d\ensuremath{\varepsilon} \ensuremath{\eta}) and agrees with finite differences to three digits. Table S9b gives, per range of \textbar G\textbar/2\(k_{n}\), the share of the sensitivity \textbar d\ensuremath{\Delta}/d\(\hat{\theta}\)\textbar{}\(^{2}\), the share of the Fourier power \textbar{}\(\hat{\theta}\)\textbar{}\(^{2}\), and the working budget: the change of the gap, in gaps, if every peak of the range grows by 100 \% at its own phase (net, opening, closing). The shell does the net work; the inner peaks, which include the differences of shell vectors, carry about half of the sensitivity with a quarter of the power but largely cancel.

\begin{table}[!htbp]
\renewcommand{\thetable}{S9}
\caption{Linear response of the gap to the Fourier coefficients, by shell of \textbar G\textbar/2\(k_{n}\).}
\centering\small
\begin{tabularx}{\linewidth}{l>{\raggedright\arraybackslash\hsize=1.43\hsize}X>{\raggedright\arraybackslash\hsize=1.43\hsize}X>{\raggedright\arraybackslash\hsize=1.43\hsize}X>{\raggedright\arraybackslash\hsize=0.36\hsize}X>{\raggedright\arraybackslash\hsize=0.36\hsize}X}
\hline\hline
\hfill\break & G\\ & / 2\(k_{n}\) & bcc 136, index 1.584: sensitivity / power / work (open, close) & bcc 136, index 1.654 & fcc 118, index 1.677 \\
\hline
0--0.5 & 0.25 / 0.09 / \ensuremath{-}0.3 (+1.0, \ensuremath{-}1.4) & 0.23 / 0.09 / \ensuremath{-}0.3 (+1.8, \ensuremath{-}2.0) & 0.32 / 0.07 / 0.0 (+1.9, \ensuremath{-}1.9) &  &  \\
0.5--0.85 & 0.27 / 0.17 / +0.7 (+3.4, \ensuremath{-}2.8) & 0.34 / 0.19 / +1.6 (+3.6, \ensuremath{-}2.0) & 0.29 / 0.22 / +2.1 (+4.7, \ensuremath{-}2.6) &  &  \\
0.85--0.95 & 0.26 / 0.20 / +3.9 (+5.2, \ensuremath{-}1.3) & 0.19 / 0.20 / +3.7 (+4.7, \ensuremath{-}1.0) & 0.18 / 0.27 / +4.7 (+6.1, \ensuremath{-}1.4) &  &  \\
0.95--1.05 & 0.17 / 0.33 / +4.3 (+5.7, \ensuremath{-}1.4) & 0.17 / 0.30 / +3.1 (+4.2, \ensuremath{-}1.1) & 0.15 / 0.21 / +4.4 (+5.3, \ensuremath{-}0.9) &  &  \\
1.05--1.15 & 0.02 / 0.02 / +0.2 & 0.03 / 0.03 / +0.1 & 0.03 / 0.04 / +0.3 &  &  \\
above 1.15 & 0.03 / 0.20 / +0.4 & 0.04 / 0.20 / +0.2 & 0.03 / 0.19 / +0.3 &  &  \\
\hline\hline
\end{tabularx}
\end{table}

\textbf{4. Lattice.} The peaks must be lattice vectors within relative thickness \ensuremath{\delta} of the sphere. A unit-volume reciprocal cell gives about 4\ensuremath{\pi}R\(^{3}\)\ensuremath{\delta} such vectors, or 2\ensuremath{\pi}R\(^{3}\)\ensuremath{\delta} pairs. Setting this equal to N with \ensuremath{\delta} = 1.2/N:

\begin{equation}
N \approx \sqrt{2.4\pi}\;R^{3/2} \approx 2.8\,R^{3/2}, \qquad b \propto N^{-1/2} \propto R^{-3/4} \propto n^{-1/4}
\end{equation}

since R = 2\(k_{n}\) \ensuremath{\propto} n\(^{1/3}\). The same argument in 2D gives c \ensuremath{-} 1 \ensuremath{\propto} n\(^{\ensuremath{-}1/2}\), the scaling of the 2D records ((c \ensuremath{-} 1)\ensuremath{\surd}n \ensuremath{\approx} 1.2--1.9).

\textbf{Against the data.} At fcc 118 (R = 4.83) the count gives N \ensuremath{\approx} 29 pairs; the designs carry 14--21 components above 0.1. Between R = 3.1 (fcc 30) and 5.1 (bcc 136) the best plane-wave c \ensuremath{-} 1 falls from 3.05 to 1.44 (1.39 in MPB). Larger cells do not continue the trend yet (Fig. 4 of the Letter): bcc 208 and fcc 202 (R = 5.8) reach 1.70 and 1.94. The scaling is a conjecture.

\textbf{Restricted support and free filling (test).} If a gap at low contrast needed only N well-placed shell peaks, re-optimizing with the support restricted to N shell pairs would find it. For N = 15, 25, 40, 60 and 90 well-spread pairs on bcc 136 and sc 88, and up to 40 on sc 40, each from start fillings 0.28, 0.37 and 0.46, no run opens a gap within the contrast range of the full-support designs: on bcc 136 and sc 88 a single run (90 pairs, start filling 0.28) opens one at all, at c = 4.66, and on sc 40 the 40-pair runs need c = 4.3--4.4 against 3.44 with the full support. With the full support the filling returns to 0.33--0.37 from either start, and the start at 0.28 does better (bcc 136: c = 2.74 against 3.09 from 0.46; sc 40: 3.44 against 3.55). The weak peaks off the shell, which thresholding creates at differences of shell vectors (item 2), are thus part of the gap, and the full-support optimizations converge to a narrow range of fillings. A one-line estimate gives a reason: for a thresholded Gaussian field the Bragg strength of a shell peak is proportional to F(t) = (c \ensuremath{-} 1)\ensuremath{\varphi}(t)/{[}1 + (c \ensuremath{-} 1)(1 \ensuremath{-} \ensuremath{\Phi}(t)){]}, whose maximum, at t = F(t), lies at filling f* = 0.364, 0.357, 0.345 and 0.341 for c = 2.3925, 2.5084, 2.7343 and 2.8108. Over 24 optimized designs the observed filling exceeds this estimate by 0.024 on average (rms difference 0.026, correlation 0.75); the design at index 1.547 has 0.372 against 0.364. The estimate ignores polarization and many-wave coupling, so it explains the range, not the optimum.

\textbf{First-order counts are not design rules.} Eq. 5 of \cite{Maiwald2022}, N = 2.46 (\(\bar{n}\)/\ensuremath{\Delta}n)\(^{2}\) gratings with a gap of 0.49 (\ensuremath{\Delta}n/\(\bar{n}\))\(^{2}\), maximizes \ensuremath{\Delta}\ensuremath{\omega}/\ensuremath{\omega} = 1.54 (\ensuremath{\Delta}n/\(\bar{n}\))/\ensuremath{\surd}N \ensuremath{-} 1.21/N: first-order half-openings at the face centres and corners of an effective zone of equal hexagons, against the rise of the Bragg frequency from face centre to corner. At the maximum both terms are of order (\ensuremath{\Delta}n/\(\bar{n}\))\(^{2}\), the order of what is left out: the coupling of a face-centre state to its six neighbouring gratings alone moves that edge by 0.26 (\ensuremath{\Delta}n/\(\bar{n}\))\(^{2}\) for any N, and their Eq. 6 allows 2 (\ensuremath{\Delta}n/\(\bar{n}\))\(^{2}\) between the polarizations, four times the predicted gap. Their 3D simulations used one N per contrast. Our count (1--4 above) rests on the same balance. Three tests. (i) Eq. 5 gives 3.0 \% at index 1.654, six times the verified 0.48 \%, and 26--46 gratings at the contrasts where 22 optimized designs lose their gaps; the effective number of shell pairs of those designs (participation number of the pairs with \textbar G\textbar/2\(k_{n}\) = 0.85--1.12) follows R, from 18 to 115 and about 0.6 R\^{}2.9 (correlation of the logarithms 0.94), and not the Eq. 5 number (0.15). (ii) Their recipe in a periodic cell: 36 equal gratings with random phases on the sphere h\(^{2}\) + k\(^{2}\) + l\(^{2}\) = 26 of an sc cell at contrast 3 (index 1.73), where Eq. 5 puts the optimum at 34 gratings and a gap of +3.5 \%. The computed bands overlap by 4.7--5.2 \% over eight phase sets and fillings 0.35 and 0.5 (cutoff 1.4--2.0 \(k_{n}\)); the same reasoning with the actual 25\textdegree{} holes of this set gives \ensuremath{-}1.7 \%, so the first-order estimate is about 3 points too optimistic. (iii) Truncation of the design at index 1.654 (313 pairs, participation number 123; amplitudes and phases kept, threshold reset; contrast 3, cutoff 1.4 \(k_{n}\)): all 313 pairs +0.25 \%, the 250 strongest (98.4 \% of the power) \ensuremath{-}0.10 \%, the 45 strongest \ensuremath{-}1.25 \%. These are one lattice set and truncations without re-optimization; the re-optimized series with the support restricted to N pairs (below) opens no gap in this contrast range with up to 90 pairs.

\section*{S10. Data and code availability}

Every claimed design can be rebuilt and re-verified from its coefficient file with one command. The best design on each cubic lattice (Letter Table I: bcc 136 at index 1.547, fcc 118 at 1.677, sc 88 at 1.681), with their structures on 80\(^{3}\) grids, their MPB claim files and band diagrams, and scripts that rebuild the structures and check the gaps with MPB, form a data package that, like the design and optimization codes and the other files listed below, is available from the author on request.

\begin{table}[!htbp]
\renewcommand{\thetable}{S10}
\caption{Data and code.}
\centering\small
\begin{tabularx}{\linewidth}{>{\raggedright\arraybackslash\hsize=0.73\hsize}X>{\raggedright\arraybackslash\hsize=1.49\hsize}X>{\raggedright\arraybackslash\hsize=0.79\hsize}X}
\hline\hline
Item & Content & Where (now) \\
\hline
Design files (one JSON per case) & lattice, n, support vectors \(m_{j}\), coefficients \(c_{j}\), threshold t, filling, \(c_{\mathrm{dense}}\) / \(c_{\mathrm{open}}\) / \(c_{03}\), extremal k-set, round history & cluster: bench/\textless campaign\textgreater/cases; best designs on branches tcm/\textless campaign\textgreater{} \\
Claim files (CLAIM3\_*.json) & entry ladder; all bands at every scan k-point; every Nelder--Mead evaluation; edge set and bands n \ensuremath{-} 2 \ldots{} n + 3 at each resolution; extrapolations & branches tcm/\texttt{p3d\_mpb3d\_claim}\_* \\
Opening files (OPEN3\_*.json) & gap at each contrast step on the edge set & branch tcm/\texttt{p3d\_mpb3d\_open} \\
Optimization histories (.npz per case) & 237 786 iterates; 84 092 binary designs with bands n \ensuremath{-} 4 \ldots{} n + 5 on kgrid(4); in the later campaigns also every scan, Nelder--Mead and contrast-search evaluation with bands n \ensuremath{-} 2 \ldots{} n + 3; format in DATA\_FORMAT.md; about 1.1 GB & cluster: bench/\textless campaign\textgreater/data \\
Code & plane-wave solver and optimizer (gopt3d.py, \texttt{gopt3d\_run.py}), dense band-edge search (densek.py), MPB claim and opening (\texttt{mpb3d\_proof.py}, \texttt{mpb3d\_open.py}), global edge search and probe (\texttt{mpb3d\_verify.py}), covering certificate (certify3d.py), band structures (\texttt{mpb3d\_bands.py}), polyhedral starts (poly\_start.py), connectivity (conn3d.py), bound (bounds3d.py), Riesz analysis (interp/peaks\_riesz.py), campaign tables and launchers & repository wave-datasets, branch p3d/go3d, benchmarks/gapdesign3d \\
\hline\hline
\end{tabularx}
\end{table}

\textbf{Re-verifying a claim.} For the design at index 1.547 (one node, about 13 h on 32 cores), with the design file of the data package; the contrast passed to the script is the plane-wave \(c_{03}\), and the first ladder rung, 0.98 \(c_{03}\) = 2.3925, is the claim contrast:

\begin{Verbatim}[breaklines=true,breakanywhere=true,fontsize=\footnotesize]
python3 mpb3d_proof.py \
  --design bcc136_index1547_design.json --c 2.44129 \
  --kextra bcc136_index1547_design.json \
  --ladder 0.98,1.02,1.06,1.1,1.15,1.2 --ladder-target 0.001 \
  --grids 6,8 --nworst 12 --res-scan 40 --nedge 10 --maxfev 40 \
  --res-final 48,64,80 --tol 1e-8 --extra 10 \
  --out OUT --name CLAIM3_bcc136_index1547
\end{Verbatim}

\textbf{Software.} MPB 1.12.0 (libctl 4.7.1) via the Meep Python interface 1.35.0; Python 3.14.8 with numpy 2.5.4 and scipy 1.18.1.

\section*{S11. Decay inside the gap, connectivity and a printed sample}

\textbf{Complex bands.} In the fixed-set plane-wave model (S2) the matrix is quadratic in k. For a real wavevector k\(_{0}\) along a surface with normal \(\hat{n}\) the evanescent waves solve det{[}M(k\(_{0}\) + \ensuremath{\kappa} \(\hat{n}\)) \ensuremath{-} \ensuremath{\nu}\(^{2}\){]} = 0 for complex \ensuremath{\kappa} (shift-invert Arnoldi), and the slowest one decays as exp(\ensuremath{-}2\ensuremath{\pi} Im \ensuremath{\kappa} x/a), with a the length of the unit-volume cell. For bcc 136 at index 1.654, over 15 channels (normals (110) and (100) and the principal axes at both band edges), Im \ensuremath{\kappa} at mid-gap is at least 0.045 and 0.062 in the median: 3.1 dB per conventional cubic cell of 2.5 \ensuremath{\lambda}\(_{0}\) for the slowest channel, 1.2 dB per vacuum wavelength. The two-band estimate \ensuremath{\lambda}\(_{0}\)/(\ensuremath{\pi} \(n_{\mathrm{eff}}\) \ensuremath{\Delta}) gives Im \ensuremath{\kappa} = 0.0047, ten times less.

\textbf{Flat gap bands, checked in MPB.} The Hessians of the MPB band-edge polishing (S5, resolution 40) give median principal curvatures of 0.47, 1.4 and 3.5 at the top of band n (11 edge points) and 0.55, 1.1 and 2.8 at the bottom of band n + 1 (15 points) (\ensuremath{\nu} per (2\ensuremath{\pi}/a)\(^{2}\)), against h = 2v\(^{2}\)/\ensuremath{\Delta}\ensuremath{\nu} \ensuremath{\approx} 140 in a two-band model with the free-photon group velocity v = c/\(n_{\mathrm{eff}}\). With the effective mass, Im \ensuremath{\kappa} = (\ensuremath{\Delta}\ensuremath{\nu}/h)\^{}\{1/2\} at mid-gap; along the steepest axis this gives 0.040--0.059 for the design at 1.654 (2.8--4.1 dB per cubic cell, slowest at the top of band n), in line with the complex bands, and 0.036--0.044 for the design at 1.584 (2.5--3.0 dB per cubic cell) (edge\_curv3d.py).

\textbf{Loss.} Flat bands also mean slow light, so absorption beside the gap is about ten times that of a homogeneous medium. Decay per cubic cell in the slowest channel of the plane-wave model of the design at 1.654, loss as a complex frequency with an electric-energy fraction of 0.6:

\begin{table}[!htbp]
\renewcommand{\thetable}{S11}
\caption{Attenuation per cubic cell beside the gap, without and with absorption.}
\centering\small
\begin{tabular}{lccc}
\hline\hline
Position & No loss & Loss tangent 0.003 & Loss tangent 0.01 \\
\hline
1 gap width below the top edge & 0 & 1.1 dB & 2.4 dB \\
top edge & 0 & 1.6 dB & 3.0 dB \\
25 \% into the gap & 2.3 dB & 2.4 dB & 3.3 dB \\
mid-gap & 3.1 dB & 3.2 dB & 3.7 dB \\
\hline\hline
\end{tabular}
\end{table}

The stop band stands 3.1 dB per cell above its surroundings without loss, 2.1 dB with loss tangent 0.003 and 1.3 dB with 0.01. Ten cubic cells give 30 dB at mid-gap without loss; at 60 GHz (cell 12.6 mm, struts 0.75--1.75 mm, air channels 1.0--2.5 mm) that is a block of 13 cm.

\textbf{Connectivity} (conn3d.py, 128\(^{3}\) voxels, periodic components with face neighbours). bcc 136 at 1.654 and fcc 118 at 1.677: both phases are single networks spanning all three directions. bcc 136 at 1.584: the dielectric has one detached grain per cell (0.42 \% of the volume), so it cannot be printed as it is. The polyhedral-start bcc 136 designs at 1.565 and 1.547: both phases are single networks spanning all three directions (Fig. 1a of the Letter).

\textbf{Tolerance.} For the design at 1.654, the gap centre moves by half the gap when the filling fraction changes by 1.1 \% or the index by 0.4 \%, so the surfaces must be uniform across the sample to about 10\(^{-3}\) \ensuremath{\lambda}\(_{0}\) (5 \textmu{}m at 60 GHz); a uniform rescaling of the whole structure only shifts the gap in frequency, but shape or index errors repeated in every cell can also narrow or close it. The filament\textquotesingle s permittivity must be measured first and the design retuned to it. A guarantee that holds for any design follows from the min--max argument of S7: a cell-periodic (or supercell-periodic) error with \textbar \ensuremath{\delta}\ensuremath{\varepsilon}\textbar{} \ensuremath{\leq} \ensuremath{\rho}\ensuremath{\varepsilon} everywhere moves every band frequency by a factor between (1 + \ensuremath{\rho})\(^{\ensuremath{-}1/2}\) and (1 \ensuremath{-} \ensuremath{\rho})\(^{\ensuremath{-}1/2}\), so a gap of relative width \ensuremath{\Delta} survives for \ensuremath{\rho} \textless{} \ensuremath{\Delta}/(1 + \ensuremath{\Delta}\(^{2}\)/4), about 0.3 \% of \ensuremath{\varepsilon} for the designs here. A displaced interface is not a small pointwise change of \ensuremath{\varepsilon} and is not covered.

\textbf{Limits.} These are decay rates of one design in the plane-wave model and effective masses from MPB, not transmission: coupling into the crystal and the diffraction orders of a 2.5 \ensuremath{\lambda}\(_{0}\) cell are not included. A finite-slab transmission calculation (ten cells, Bloch-periodic sides) remains to be done, as does the same analysis for the design at 1.547, whose gap is about 0.6 times that of the design at 1.654.

\section*{S12. Polyhedral starts}

\textbf{Why a polyhedron.} The optimizer starts from a small set of strong Fourier peaks and then lets every coefficient of the support move (S3). A start that already puts the peaks where the Bragg planes of the gap belong spares it the hardest part of the work. If the peaks sit at the vertices of a polyhedron inscribed in the Bragg sphere \textbar G\textbar{} = 2\(k_{n}\), their Bragg planes k\ensuremath{\cdot}G = \textbar G\textbar{}\(^{2}\)/2 bound the dual polyhedron: peaks at the vertices of an Archimedean solid make a Catalan solid the Bragg polyhedron, or Jones zone, of the start. A good start meets three conditions. The zone should be nearly round, so that it holds the free-photon states of the lowest n bands at about the same frequency in every direction; then twice its volume (two polarizations, primitive cell of volume 1) is close to n. Its vertices must lie close to reciprocal-lattice vectors. And band n should sit at a deep minimum of the empty-lattice bound (S7). Solids with icosahedral symmetry have the roundest zones, but no lattice carries five-fold symmetry, so their vertices need rational approximants of the golden ratio \ensuremath{\varphi} = (1 + \ensuremath{\surd}5)/2.

\textbf{The icosidodecahedral start.} The icosidodecahedron is the Archimedean solid whose 30 vertices are the edge midpoints of an icosahedron; its Catalan dual is the rhombic triacontahedron, with 30 rhombic faces. With circumradius \ensuremath{\varphi} its vertices are

\begin{itemize}
\tightlist
\item
  the 6 points (\ensuremath{\pm}\ensuremath{\varphi}, 0, 0), (0, \ensuremath{\pm}\ensuremath{\varphi}, 0), (0, 0, \ensuremath{\pm}\ensuremath{\varphi});
\item
  the 24 points (\ensuremath{\pm}\ensuremath{\varphi}\(^{2}\)/2, \ensuremath{\pm}\ensuremath{\varphi}/2, \ensuremath{\pm}1/2), (\ensuremath{\pm}\ensuremath{\varphi}/2, \ensuremath{\pm}1/2, \ensuremath{\pm}\ensuremath{\varphi}\(^{2}\)/2), (\ensuremath{\pm}1/2, \ensuremath{\pm}\ensuremath{\varphi}\(^{2}\)/2, \ensuremath{\pm}\ensuremath{\varphi}/2), with all sign combinations: the cyclic permutations of one triple.
\end{itemize}

They form 15 \ensuremath{\pm}G pairs. Scaled by 4, they become (\ensuremath{\pm}4\ensuremath{\varphi}, 0, 0) \ensuremath{\approx} (\ensuremath{\pm}6.47, 0, 0) and the cyclic permutations of (\ensuremath{\pm}2\ensuremath{\varphi}\(^{2}\), \ensuremath{\pm}2\ensuremath{\varphi}, \ensuremath{\pm}2) \ensuremath{\approx} (\ensuremath{\pm}5.24, \ensuremath{\pm}3.24, \ensuremath{\pm}2). The nearest reciprocal-lattice vectors of bcc (conventional Miller indices with h + k + l even) are the orbit \{600\} and the cyclic orbit \{532\}c, that is (5, 3, 2), (3, 2, 5) and (2, 5, 3) with all signs. Since 5, 3 and 2 are consecutive Fibonacci numbers, this approximant replaces \ensuremath{\varphi} by 5/3 and 3/2; the next one, 8/5, is \{10 0 0\} + \{853\}c on bcc 576. The two orbits have radii 6 and \ensuremath{\surd}38, 2.7 \% apart; the six axial vectors point exactly along exact vertices, the other 24 within 1.2\textdegree{} (Fig. S2a). For bcc 136 they lie at 0.94 and 0.97 \ensuremath{\times} 2\(k_{n}\). Their 30 Bragg planes bound an approximate rhombic triacontahedron (Fig. S2b) whose volume, doubled for the two polarizations, is exactly 2038/15 = 135.87: the zone holds almost exactly the 136 bands below the gap. This match, more than the Archimedean shape itself, is why the start fits bcc 136. Translated by the lattice, the zone holds exactly 68 vectors k + G at 93.6 \% of all Bloch vectors k, 67 at 6.2 \% and 66 at 0.2 \%, never more (200 000 random k; rms deviation from 68: 0.27); the 8/5 approximant holds exactly 288 at 97.1 \% (rms 0.18). No other start comes close (Table S12a). The same count gives 191 for the truncated dodecahedron on bcc 190, but 83--145 for the other sets, which a bare-structure screen chose instead (Table S12a). The cyclic orbit \{532\}c and its mirror image \{523\}c are the two icosidodecahedra compatible with the cubic axes; the icosahedron whose edge midpoints form \{532\}c is \{350\}c, and its vectors are the five-fold axes of the start.

\textbf{Fibonacci volume identity.} The construction continues with consecutive Fibonacci numbers r \textless{} q \textless{} p (2, 3, 5 above). On bcc the 30 vectors \{2q 0 0\} + \{pqr\}c bound a zone whose doubled volume N, in units of a primitive cell of volume 1, is exact:

\begin{equation}
\begin{aligned}
N_m &= 4F_{3m-3} - \frac{2F_{m-3}}{F_m F_{m+1}}, \qquad (r, q, p) = (F_{m-1}, F_m, F_{m+1}),\\
N &= 20r^3 - 12\delta r + \frac{2(q-2r)\,\delta^2}{q(q+r)}, \qquad \delta = r^2 + qr - q^2 = (-1)^m,\\
F_{3j} &= 5F_j^3 + 3(-1)^j F_j .
\end{aligned}
\end{equation}

For m = 4, 5, 6 and 7 the first line gives 136 \ensuremath{-} 2/15, 576 \ensuremath{-} 1/20, 2440 \ensuremath{-} 1/26 and 10336 \ensuremath{-} 2/91; we checked it numerically up to m = 8. Proof: in Miller-index units the zone volume equals N. Its 30 half-spaces have 32 vertices, (\ensuremath{\pm}d, \ensuremath{\pm}d, \ensuremath{\pm}d) with d = (q\(^{2}\) + qr + r\(^{2}\))/{[}2(q + r){]}, and the signed cyclic permutations of (0, q, r\(^{2}\)/q) and (0, r, q). Summing the pyramids over the 6 axial and 24 oblique facets gives the second line, with \textbar \ensuremath{\delta}\textbar{} = 1 by Cassini\textquotesingle s identity, and the tripling identity (third line, j = m \ensuremath{-} 1) turns 20r\(^{3}\) \ensuremath{-} 12\ensuremath{\delta}r into the integer of the first line. Each candidate band is four times the previous one plus the one before (2440 = 4 \ensuremath{\times} 576 + 136). The two orbits differ by 2 in squared Miller radius, so their relative radial spread, about 1/(4q\(^{2}\)), falls from 2.7 \% at band 136 to 1.0 \% at band 576 and keeps falling; the angular errors fall too (1.2\textdegree{} and 0.5\textdegree{}, Table S12b). A closer approximant does not by itself open a gap or lower the contrast: the bare 8/5 start still has none at contrast 4.5.

\textbf{What the optimizer keeps.} The bare start has no gap: at contrast 3 its bands n and n + 1 overlap by about 1.3 \%, against 2.4 \% for the Riesz-weighted shell. Imposing the start\textquotesingle s symmetry during the optimization stalls near zero gap, so only the start is polyhedral, and every coefficient is free afterwards. The final designs keep the polyhedron: in the design at index 1.547 the strongest shell peak and 26 of the 40 strongest are start vectors (Fig. 1b), and the 30 start vectors hold 36 \% of the Fourier power in the shell 0.85--1.12 \ensuremath{\times} 2\(k_{n}\). The next 9.6 \% sits in the 12 vectors \{602\}c at 0.991 \ensuremath{\times} 2\(k_{n}\) (median amplitude 51 \% of the strongest peak), one in each of the twelve five-fold holes, 12.5\textdegree{} from its five-fold axis and 18.4\textdegree{} from the nearest start vector. The same orbit holds 11.1 \% of the shell power in the design at index 1.565 and 8.2 \% in the pushed IDIC design. Even a design grown from a random start puts 15 \% of its shell power on the 30 start vectors and 4.7 \% on \{602\}c, which are 6.7 \% and 2.7 \% of the 446 shell vectors. Apart from \{602\}c, the peaks of the design at index 1.547 avoid the holes: they put 12 \% of their power within 15\textdegree{} of the five-fold axes, where 29 \% of the shell vectors lie. The holes are thus covered by one peak each, at the sphere, not by many weak ones. IDIC offers an icosahedron \{350\}c at the hole centres, but inside the sphere at 0.914 \ensuremath{\times} 2\(k_{n}\); the optimizer removes it (0.9 \% of the power, median amplitude 16 \%) and grows \{602\}c instead. The other families keep part of their start (Fig. 4, insets): A keeps the octahedron \{400\} (median amplitude 97 \%) and part of \{322\} (54 \%) and adds \{411\} and \{331\} (42 \% and 35 \%) between them; TD keeps \{136\}c (59 \%) and loses \{017\}c and \{345\}c (22 \% and 25 \%).

\begin{figure}[tb]
\includegraphics[width=\linewidth]{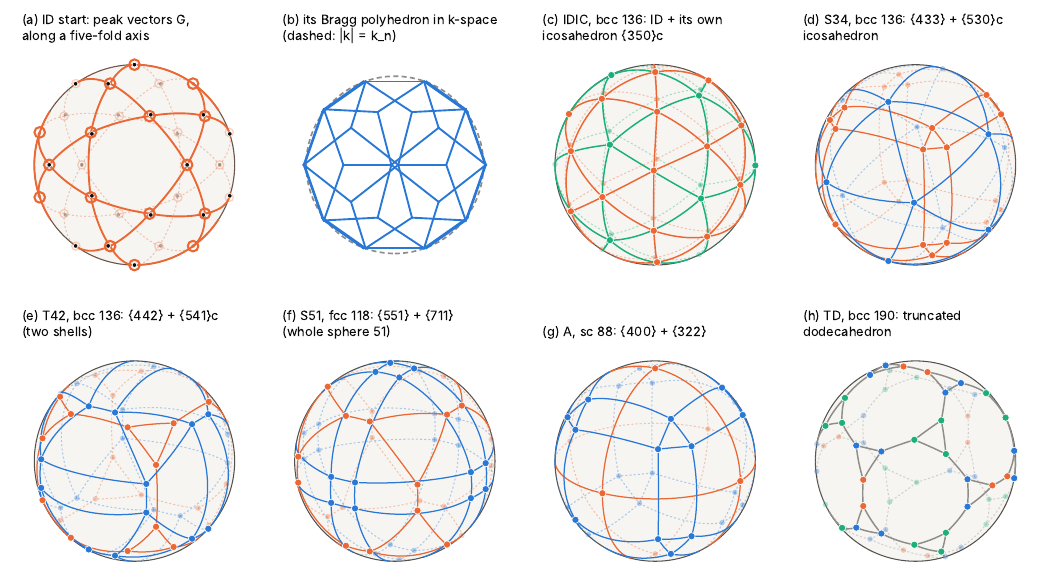}
\caption{Polyhedral starts. (a) The icosidodecahedral start of bcc 136 in reciprocal space, seen along a five-fold axis: directions of its peak vectors G, \{600\} + \{532\}c (orange rings), and of the exact icosidodecahedron (black dots), on the sphere \textbar G\textbar{} \ensuremath{\approx} 2\(k_{n}\); pale: its far half. (b) The same peaks seen by the light: the region of wavevectors k bounded by their 30 Bragg planes k\ensuremath{\cdot}G = \textbar G\textbar{}\(^{2}\)/2, each plane bisecting a peak vector, drawn at the scale of k (half that of G) from the same direction. It is an approximate rhombic triacontahedron, the Catalan dual of the icosidodecahedron, with its faces at 0.94--0.97 \(k_{n}\), just inside the free-photon sphere \textbar k\textbar{} = \(k_{n}\) of band n (dashed). A plane wave with k inside it is not Bragg-reflected by the start peaks, the gap opens on its faces, and twice its volume counts the bands below the gap. (c)--(h) The start sets of the other families by direction; colours mark the orbits, each drawn with its own edges (grey: the truncated dodecahedron).}
\label{fig:S2}
\end{figure}

\textbf{Phases.} The design at index 1.547 is not centrosymmetric: no choice of origin makes its coefficients real (the best origin leaves a power-weighted rms phase of 43\textdegree{} off 0 or \ensuremath{\pi}, against 45\textdegree{} for random phases). In the frame of the start, the strong vertex peaks keep its signs: \{600\} at phases 5--19\textdegree{} and \{532\}c at 129--176\textdegree{} (Fig. S3a). The weaker peaks, \{602\}c included, have phases spread as if at random. The phases matter as much as the amplitudes (Fig. S3b). On 54 k-points at the plane-wave \(c_{03}\) = 2.441, where the design has a gap of 0.65 \%, a random jitter of all phases by 5\textdegree{} costs little, 10\textdegree{} halves the gap and 20\textdegree{} closes it, with the amplitudes and the filling held fixed. Random phases for the 30 strongest pairs, or for the other 280, turn the gap into an overlap of 0.9--1.6 \%, and setting the 30 strongest to exactly 0 or \ensuremath{\pi} gives an overlap of 0.65 \%. The gap rests on phase relations among many peaks, the weak ones included, not on the vertex amplitudes alone.

\begin{figure}[tb]
\includegraphics[width=\linewidth]{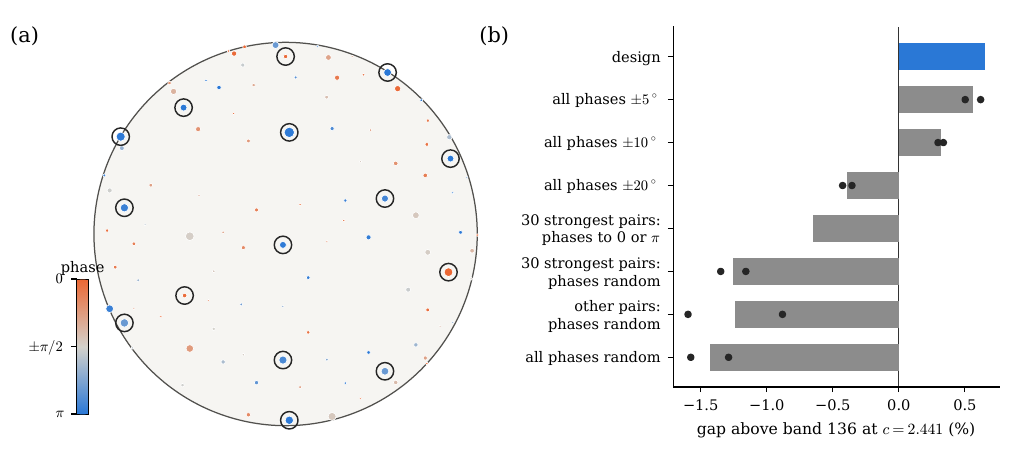}
\caption{Phases of the Fourier peaks of the design at index 1.547. (a) Peaks \(\hat{\theta}\)(G) of the binary structure in the shell 0.85--1.12 \ensuremath{\times} 2\(k_{n}\) by direction on the Bragg sphere, near half (area \ensuremath{\propto} \textbar{}\(\hat{\theta}\)\textbar{}\(^{2}\), as in Fig. 1b), coloured by the cosine of the phase, with the origin of the start set: orange, phase 0; blue, \ensuremath{\pi}; grey, \ensuremath{\pm}\ensuremath{\pi}/2. Rings: the 30 start vectors, \{600\} (sign +) and \{532\}c (sign \ensuremath{-}). (b) Gap above band 136 at c = 2.441 (the plane-wave \(c_{03}\)) on 54 k-points, for the design and for versions with modified coefficient phases; amplitudes and filling (0.372) held fixed. Dots: two random seeds; bars: their mean.}
\label{fig:S3}
\end{figure}

\textbf{Families tested.} Table S12a lists every polyhedral family, with the plane-wave contrast of its best start before and after one round of perturbed restarts, and its MPB claims. Apart from ID, IDIC and TD, the sets are not named solids but compounds of two cubic orbits, each an exact polyhedron on the sphere: an orbit \{hhk\} has the faces of a truncated cube (TC-type), and an orbit \{hkk\} or a cyclic orbit \{hkl\}c those of a rhombicuboctahedron (RCO-type), deformed unless the ratios are the Archimedean ones (Fig. S2d--g); T42 has its two orbits at radii 6 and \ensuremath{\surd}42, S51 fills the whole sphere h\(^{2}\) + k\(^{2}\) + l\(^{2}\) = 51. They are the best sets of a bare-structure screen of all single orbits and orbit pairs of the shell 0.88--1.04 \ensuremath{\times} 2\(k_{n}\) (105 sets on bcc 136, 55 on fcc 118 and 45 on sc 88), chosen by that screen and not by a geometric rule. S34 holds 36 of the 48 vectors with h\(^{2}\) + k\(^{2}\) + l\(^{2}\) = 34, one of the two icosahedra on that sphere being absent, and the outer shell of T42 lies just outside the Bragg sphere. IDIC is the icosidodecahedron with its own icosahedron, \{600\} + \{532\}c + \{350\}c, run with random signs, IDIC (random), and with symmetric signs, IDIC (symmetric); the compound with the mirrored icosahedron was not run. Its zone holds 124 states instead of 136, and its best start, IDIC (symmetric), reaches c = 2.70 in plane waves (index 1.64), against 2.50 for the icosidodecahedron alone, so adding its icosahedron does not help; one round of perturbed restarts brings it to c = 2.60 (index 1.61), against 2.44 after two rounds of the icosidodecahedron alone, and MPB confirms IDIC (symmetric) at index 1.628 unpushed and at 1.594 after one round, and IDIC (random) at 1.614 after one round (S6). The pushed claims at 1.547 (bcc 136) and 1.681 (sc 88) pass the global edge search, and their plane-wave models carry numerical covering certificates.

\begin{table}[!htbp]
\renewcommand{\thetable}{S12a}
\caption{Polyhedral families: start sets, Bragg zones, best plane-wave contrasts and MPB claims. Exactly n/2: share of Bloch vectors k at which the zone holds exactly n/2 vectors k + G (200 000 random k for ID, IDIC, A, TD and ID 8/5; 4096 scrambled Sobol points otherwise), with the rms deviation of the count from n/2. IDIC: one start set, run with random and with symmetric signs.}
\centering\footnotesize
\begin{tabularx}{\linewidth}{lc>{\raggedright\arraybackslash\hsize=1.11\hsize}X>{\raggedright\arraybackslash\hsize=1.11\hsize}Xcc>{\raggedright\arraybackslash\hsize=0.58\hsize}X>{\raggedright\arraybackslash\hsize=1.11\hsize}X>{\raggedright\arraybackslash\hsize=1.11\hsize}X}
\hline\hline
Family & Cell & Orbits (\ensuremath{\pm}G pairs) & Shape & Radius / 2\(k_{n}\) & 2 \ensuremath{\times} zone volume & Exactly n/2 (rms) & Best c, start / pushed & MPB claims (index) \\
\hline
ID & bcc 136 & \{600\} + \{532\}c (15) & icosidodecahedron (5/3) & 0.94--0.97 & 135.87 & 94 \% (0.27) & 2.50 / 2.44 (two rounds) & 1.572; 1.565 and 1.555 (1 round); 1.547 (2 rounds) \\
IDIC & bcc 136 & \{600\} + \{532\}c + \{350\}c (21) & ID + its own icosahedron & 0.91--0.97 & 123.53 & 0 \% (6.3) & random signs 2.86 / 2.66; symmetric 2.70 / 2.60 & 1.614 (random, 1 round; unpushed: candidate at 1.675); 1.628 (symmetric); 1.594 (symmetric, 1 round) \\
S34 & bcc 136 & \{433\} + \{350\}c (18) & RCO-type + icosahedron & 0.91 & 122.08 & 0 \% (7.6) & 2.86 / 2.82 & 1.675; 1.661 (1 round) \\
T42 & bcc 136 & \{442\} + \{541\}c (24) & TC-type + RCO-type & 0.94--1.02 & 144.45 & 17 \% (6.5) & 3.08 / 2.78 & 1.649 (1 round) \\
S51 & fcc 118 & \{551\} + \{711\} (24) & TC-type + RCO-type & 0.93 & 104.80 & 6 \% (7.1) & 3.01 / 2.83 & candidates only (1.718; 1.666 after 1 round) \\
A & sc 88 & \{400\} + \{322\} (15) & octahedron + RCO-type & 0.91--0.94 & 83.16 & 28 \% (3.2) & 2.91 / 2.88 & 1.687; 1.681 (1 round) \\
TD & bcc 190 & \{017\} + \{136\} + \{345\} (30) & truncated dodecahedron & 0.95--0.99 & 191.02 & 18 \% (2.5) & 3.04 / --- & candidate only (1.762) \\
ID 8/5 & bcc 576 & \{10 0 0\} + \{853\}c (15) & icosidodecahedron (8/5) & 0.96--0.97 & 575.95 & 97 \% (0.18) & not optimized & --- \\
\hline\hline
\end{tabularx}
\end{table}

For comparison, the Riesz-weighted shell gives c = 2.79 on bcc 136 (two rounds: 2.56, claim at index 1.584), 2.95 on fcc 118 and 3.14 on sc 88 with the same settings.

\textbf{Archimedean and Catalan families.} Table S12b lists the icosahedral Archimedean solids, and the vertex directions of the Catalan solids (the symmetry axes), with their best lattice approximants up to R = 8.6, where R is the shell radius in units of a primitive cell of volume 1. The cubic Archimedean solids need no approximant: every orbit \{hkl\} of the cubic group is an exact polyhedron on the lattice (the cuboctahedron \{110\}, the truncated octahedron \{210\}), but their zones are less round. Among the icosahedral solids, only the icosidodecahedron has a close approximant at a radius where the empty-lattice bound is low.

\begin{table}[!htbp]
\renewcommand{\thetable}{S12b}
\caption{Lattice approximants of the icosahedral Archimedean solids and of the Catalan vertex directions, up to R = 8.6.}
\centering\small
\begin{tabularx}{\linewidth}{>{\raggedright\arraybackslash\hsize=0.72\hsize}Xc>{\raggedright\arraybackslash\hsize=1.28\hsize}X}
\hline\hline
Solid & \ensuremath{\pm}G pairs & Best approximants up to R = 8.6 (radial spread, direction error) \\
\hline
Icosahedron & 6 & sc \{320\}c, R 3.6, n 64 (2.0\textdegree{}); bcc \{530\}c, R 4.6, n 136 (0.8\textdegree{}) \\
Dodecahedron & 10 & sc \{333\} + \{520\}c, R 5.3, n 190 (3.6 \%, 0.9\textdegree{}); sc \{555\} + \{830\}c, R 8.6, n 800 (1.4 \%) \\
Icosidodecahedron & 15 & bcc \{600\} + \{532\}c, R 4.8, n 136 (2.7 \%, 1.2\textdegree{}); bcc \{10 0 0\} + \{853\}c, R 7.9, n 576 (1.0 \%, 0.5\textdegree{}) \\
Truncated icosahedron & 30 & none below R = 8; bcc R 8.1, n 590; sc R 8.2, n 602 \\
Truncated dodecahedron & 30 & bcc R 5.5, n 190 (4.1 \%); bcc R 6.5, n 316 (1.5 \%, 4.3\textdegree{}) \\
Rhombicosidodecahedron & 30 & bcc R 6.7, n 334 (2.8 \%, 1.0\textdegree{}) \\
Truncated icosidodecahedron & 60 & none within 5 \% and 4.5\textdegree{} \\
Snub dodecahedron with its mirror image & 60 & bcc R 7.6, n 460 (4.4 \%, 3.8\textdegree{}); bcc R 8.5, n 656 (4.4 \%, 3.1\textdegree{}) \\
Catalan vertex directions (symmetry axes) & 7, 13, 16, 31 & cubic axes: sc \{700\} + \{444\}, R 7.0 (1.0 \%); icosahedral 16 and 31: nothing within 4 \% \\
\hline\hline
\end{tabularx}
\end{table}

Bare structures of the larger sets (n = 190 to 656) open no gap at band n even at contrast 4.5 (closest: the 8/5 icosidodecahedron on bcc 576, still overlapping by 0.3 \%), and where checked, band n is not singled out among its neighbours, unlike band 136 of bcc; none was optimized except the truncated dodecahedron on bcc 190. First optimizations of further starts end at higher contrast in plane waves, without push rounds: the rhombicosidodecahedron on bcc 334 at c = 2.93 (index 1.71), the rhombicuboctahedron on fcc 165 at 3.10 (1.76), and the icosidodecahedral start of bcc 136 with an added cylindrical-wave background at 2.57 (1.60), against 2.50 without it. The 8/5 icosidodecahedron on bcc 576 is still being optimized.

\section*{S13. Two-dimensional designs}

\textbf{Setting.} TM polarization in square and triangular cells of unit area. A gap counts when \ensuremath{\Delta}\ensuremath{\omega}/\(\ensuremath{\omega}_{\mathrm{mid}}\) above band n reaches 0.3 \%. With \ensuremath{\pi}\(\ensuremath{\rho}_{n}^{2}\) = n, the Bragg circle of band n is \textbar G\textbar{} = 2\(\ensuremath{\rho}_{n}\), and the empty-lattice bound \(c_{B}\)(n) follows from the min--max argument of S7.

\textbf{Equal-rod records.} N equal circular rods (N = n, except 37 rods at square 41) start from eight families of point arrangements (optimal toric packings, Riesz-energy minimizers with s = 1, 2, 6 and 12, maxmin arrangements, symmetric and toric centroidal Voronoi tessellations) or from the maxima of quasiperiodic interference fields. Their positions and common radius are optimized along a contrast path with a plane-wave TM solver: exact disk Fourier coefficients, Hellmann--Feynman gradients of a smoothed minimum of the gap, centre distances of at least 1.02 \ensuremath{\times} 2r, and a step accepted when the refined plane-wave gap reaches 0.33 \%. The records of Table S13 are a quasiperiodic-field start (square 41) and Riesz s = 12 starts (square 52, triangular 45 and 58); they fill 34--39 \% of the cell.

\textbf{Dense MPB check.} Every MPB value in Table S13 comes from MPB with n + 6 bands at resolution 128 (square) or 144 (triangular): a 25 \ensuremath{\times} 49 k-grid over the Brillouin zone, Nelder--Mead refinement from the three highest points of band n and the three lowest of band n + 1, and a contrast ladder (rods: 1, 1.005, 1.01, \ldots{} times the optimizer\textquotesingle s contrast; free form: steps of 0.5 \% upward from 0.995 times the plane-wave contrast). The free-form values interpolate the ladder linearly to a 0.3 \% gap.

\textbf{Free-form designs.} The level-set method of S2 and S3 with a 2D TM plane-wave solver, at square n = 28--208 and triangular n = 31--205 (114 optimization histories). At every n, 4--7 independent starts end within about 0.5 \% of each other, and a support up to 1.5 \ensuremath{\times} 2\(\ensuremath{\rho}_{n}\) instead of 1.06 \ensuremath{\times} 2\(\ensuremath{\rho}_{n}\) gains at most 0.2 \%. The designs fill 40--47 \% of the cell, rising with n.

\begin{table}[!htbp]
\renewcommand{\thetable}{S13}
\caption{Lowest contrast c for a 0.3 \% TM gap above band n in 2D, with the checks. MPB ladder: contrast and gap (\%) at the two rungs around 0.3 \%.}
\centering\small
\begin{tabularx}{\linewidth}{lc>{\raggedright\arraybackslash\hsize=0.51\hsize}X>{\raggedright\arraybackslash\hsize=1.67\hsize}X>{\raggedright\arraybackslash\hsize=0.51\hsize}Xc>{\raggedright\arraybackslash\hsize=1.31\hsize}X}
\hline\hline
Cell, n & Rods (dense MPB) & Free form, plane waves & Free form, MPB ladder & Free form, MPB at 0.3 \% & \(c_{B}\) & (c \ensuremath{-} 1)/(\(c_{B}\) \ensuremath{-} 1) \\
\hline
square 41 & 1.2776 & 1.2149 & 1.2148: 0.254; 1.2209: 0.315 & 1.219 & 1.0753 & 2.9 \\
square 52 & 1.2565 & 1.2339 & 1.2339: 0.288; 1.2400: 0.361 & 1.235 & 1.0536 & 4.4 \\
triangular 45 & 1.2494 & 1.2080 & 1.2077: 0.281; 1.2137: 0.340 & 1.2096 & 1.0676 & 3.1 \\
triangular 58 & 1.2132 & 1.1827 & 1.1827: 0.270; 1.1886: 0.336 & 1.185 & 1.0433 & 4.3 \\
triangular 180 & --- & 1.1425 & not run & --- & --- & 5.4 (plane waves) \\
\hline\hline
\end{tabularx}
\end{table}

A second triangular-45 design (plane waves 1.2077) reaches 0.3 \% at 1.208 in MPB; Table S13 keeps the more conservative value. In these four cells MPB needs 0.1--0.4 \% more contrast than the plane-wave solver for the same gap.

\textbf{Fourier peaks.} The triangular n = 12 rod record (12 rods in rhombic tetramers, c = 1.3583, filling 0.47) carries a regular 12-gon of peaks at 0.991 \ensuremath{\times} 2\(\ensuremath{\rho}_{n}\) and 6 inner peaks at 0.476 \ensuremath{\times} 2\(\ensuremath{\rho}_{n}\); a regular 12-gon of area n has its faces at 0.988 \ensuremath{\times} 2\(\ensuremath{\rho}_{n}\). Starts that put Fourier peaks on the Bragg ring alone reproduce the square rod records to within 0.1 \% (n = 20: 3 \% above) but do not beat them. In the rod records, 79 \% of the strong inner peaks (0.05 \textless{} \textbar G\textbar/2\(\ensuremath{\rho}_{n}\) \textless{} 0.85) are differences \(G_{a}\) \ensuremath{-} \(G_{b}\) of two strong ring vectors, against 57 \% for the same number of random inner vectors (29 of 37 records above random; Wilcoxon p = 8 \ensuremath{\times} 10\(^{-5}\)). In the free-form designs the share is 50--100 \%, against 11--45 \% at random. The Bragg lines of the strongest ring peaks bound a zone of area close to n in 13 of 19 free-form designs (p \ensuremath{\leq} 0.05 against a random-order null), and only weakly in the rod records.

\textbf{Data.} The rod designs, the free-form designs and every MPB ladder are kept with the 3D data (S10).

\end{document}